\documentclass[usenames,dvipsnames]{aastex63}

\usepackage{graphicx}	% Including figure files
\usepackage{amsmath}	% Advanced maths commands
\usepackage{amssymb}	% Extra maths symbols
\usepackage{relsize}
\usepackage{mathtools}
\usepackage{bigints}

\usepackage{floatrow}
\usepackage{ulem}
\usepackage{soul}
\usepackage{comment}
\usepackage{bm}

\usepackage{macros_vdb}
\graphicspath{{./}{figures/}}

\received{XXX}
\revised{YYY}
\accepted{ZZZ}

\begin{document}

\shorttitle{A Self-Consistent Treatment of Dynamical Friction}
\shortauthors{Banik and van den Bosch}

\title{A Self-Consistent, Time-Dependent Treatment of Dynamical Friction:\\ New Insights regarding Core Stalling and Dynamical Buoyancy}

\correspondingauthor{Uddipan Banik}
\email{uddipan.banik@yale.edu}

\author[0000-0002-9059-381X]{Uddipan Banik}
\affiliation{Department of Astronomy, Yale University, PO. Box 208101, New Haven, CT 06520, USA}

\author[0000-0003-3236-2068]{Frank~C.~van den Bosch}
\affiliation{Department of Astronomy, Yale University, PO. Box 208101, New Haven, CT 06520, USA}

\label{firstpage}

%---------------------------------------------------------------

\begin{abstract}
  Dynamical friction is typically regarded a secular process, in which the subject (`perturber’) evolves very slowly (secular approximation), and has been introduced to the host over a long time (adiabatic approximation). These assumptions imply that dynamical friction arises from the LBK torque with non-zero contribution only from pure resonance orbits. However, dynamical friction is only of astrophysical interest if its timescale is shorter than the age of the Universe. In this paper we therefore relax the adiabatic and secular approximations. We first derive a generalized LBK torque, which reduces to the LBK torque in the adiabatic limit, and show that it gives rise to transient oscillations due to non-resonant orbits that slowly damp out, giving way to the LBK torque. This is analogous to how a forced, damped oscillator undergoes transients before settling to a steady state, except that here the damping is due to phase mixing rather than dissipation. Next, we present a self-consistent treatment, that properly accounts for time-dependence of the perturber potential and circular frequency (memory effect), which we use to examine orbital decay in a cored galaxy. We find that the memory effect results in a phase of accelerated, super-Chandrasekhar friction before the perturber stalls at a critical radius, $\Rcrit$, in the core (core-stalling). Inside of $\Rcrit$ the torque flips sign, giving rise to dynamical buoyancy, which counteracts friction and causes the perturber to stall. This phenomenology is consistent with $N$-body simulations, but has thus far eluded proper explanation.
\end{abstract}

\keywords{
Dynamical friction ---
Orbital resonances ---
Gravitational interaction ---
Stellar dynamics ---
Galaxy dark matter halos ---
Galaxy nuclei}

%---------------------------------------------------------------

\section{Introduction}
\label{sec:intro}

Dynamical friction is an important ingredient of hierarchical structure formation. It is the dynamical process by which galaxies merge, and by which globular clusters and black holes sink to the centers of their host systems where they can form bulges and binary black holes, respectively. In his seminal 1943 paper, Chandrasekhar showed that dynamical friction arises from the transfer of energy and momentum from the subject to the individual particles that make up the host system traversed by the subject. In particular, \cite{Chandrasekhar.43} considered a subject mass $M$ moving on a straight orbit through a uniform and isotropic sea of background (or `field') particles of mass $m \ll M$. When a field particle encounters the subject, it experiences velocity changes $\Delta v_{\perp}$ and $\Delta v_{\parallel}$ in the directions perpendicular and parallel to the direction of the relative velocity. Chandrasekhar summed the velocity changes from the encounters between the subject mass and all field particles, treating them as independent two-body encounters, and showed that the net result is a frictional force acting on $M$ given by
\begin{equation}\label{aDF}
\bF_{\rm DF} = - \frac{4 \pi G^2 M^2}{v^2} \, \ln\Lambda \, \rho(<v) \, \frac{\bv}{v}\,,
\end{equation}
\citep[see e.g.,][]{MBW10}. Here $\bv$ is the velocity of the subject mass, $\rho(<v)$ is the density of field particles with a speed less than $v = \vert \bv \vert$, and $\ln\Lambda = \ln[b_{\rm max}/b_{\rm min}]$ is the Coulomb logarithm, with $b_{\rm max}$ and $b_{\rm min}$ the maximum and minimum impact parameters of the encounters contributing to the drag.

Equation~(\ref{aDF}) is used routinely in astrophysics, even though it formally only applies to a uniform, isotropic background of field particles. While numerous studies have shown that it gives a reasonably accurate description of the orbital decay rates in galaxies and dark matter halos \citep[e.g.,][]{Lin.Tremaine.83, Cora.etal.97, vdBosch.etal.99, Hashimoto.etal.03, Boylan-Kolchin.etal.08, Jiang.etal.08}, there are also cases in which it clearly fails. For instance, according to equation~(\ref{aDF}) the drag force is proportional to the local density. Hence, a subject mass orbiting outside of a galaxy or halo of finite extent should experience no drag. This is inconsistent with numerical experiments, which show that even in such cases the subject loses orbital angular momentum \citep[][]{Lin.Tremaine.83}. Another example where the standard treatment of dynamical friction fails is `core-stalling', the cessation of dynamical friction in the central constant-density core of a halo or galaxy \citep[e.g.,][]{Read.etal.06c, Inoue.11, Cole.etal.12, Petts.etal.15, Petts.etal.16, DuttaChowdhury.etal.19}.

Since the seminal work by Chandrasekhar, dynamical friction has been studied using a variety of different techniques and aproaches. This includes the Fokker-Planck method, in which dynamical friction arises from the momentum exchange described by the first-order diffusion coefficient \citep[][]{Rosenbluth.etal.57, Binney.Tremaine.08}, stochastic approaches based on the fluctuation-dissipation theorem, in which dynamical friction arises from a correlation between the perturber's velocity vector and the stochastic force it experiences from the field particles \citep[][]{Bekenstein.Maoz.92, Maoz.93, Nelson.Tremaine.99, Fouvry.Bar-Or.18}, and a wide variety of methods that treat dynamical friction as a drag force arising from a `wake', or `polarization cloud' developing behind the perturber \citep[][]{Marochnik.68, Kalnajs.71, Mulder.83, Weinberg.89, Colpi.Pallavicini.98}. An excellent in-depth account of how all these methods relate to a generalized Landau equation derived from a truncation of the BBGKY hierarchy issued from the Liouville equation can be found in \citet{Chavanis.13}.

A shortcoming of many, though not all, of these methods is that they only treat dynamical friction as a local phenomenon and/or that they have only been worked out for perturbers moving through homogeneous density distributions. The first study to overcome this, and to treat dynamical friction in a more realistic, non-uniform density distribution, is that by \citet[][hereafter TW84]{Tremaine.Weinberg.84}. Using a perturbative approach, in which the subject mass, $M,$ is treated as a small, time-dependent perturbation on a circular orbit in a spherical system, they show that dynamical friction is entirely due to resonant orbits that give rise to a net retarding torque on the perturbing subject mass\footnote{Throughout this paper we will use `subject' and `perturber' without distinction.}. In particular, by integrating the torque exerted by a single resonant field particle, multiplied by the (unperturbed) velocity distribution function along the orbits perturbed to second order and summing over all resonances, they obtain a torque that is second order in the perturber's mass (i.e., proportional to $M^2$). This torque is known as the LBK torque, after \citet{LyndenBell.Kalnajs.72} who first derived it in their treatment of angular momentum transport due to spiral structure in disk galaxies. Note that this resonance picture of dynamical friction gives a natural explanation for the non-zero drag experienced by a subject mass orbiting outside of the galaxy, as it simply arises from the net torque due to resonant interactions with stars inside of the galaxy. 

A slightly different perturbative approach was recently taken by \citet[][hereafter KS18]{Kaur.Sridhar.18}; rather than perturbing the resonance orbits, they use the collisionless Boltzmann equation to compute the perturbed distribution function of field particles, which they integrate along the unperturbed resonant orbits. This once again yields a net torque that is second order in $M,$ and consistent with the LBK torque obtained by TW84. Interestingly, KS18 then proceed to show that when the perturber enters the core region of a galaxy the number of low-order resonances (which dominate the torque) is suppressed and the strength of the remaining resonances is weakened. Hence, core stalling has a natural explanation in terms of the LBK torque. However, it fails to explain two related phenomena that have been identified in numerical simulations. The $N$-body simulations by \cite{Cole.etal.12} manifest `dynamical buoyancy' in that perturbers initially placed near the center of a cored galaxy are found to be `pushed out'. Others have reported that when a perturber approaches a core, it first experiences a phase of strongly enhanced `super-Chandrasekhar' dynamical friction, followed by a `kick-back' effect in which the pertuber is pushed out again \citep[][]{Goerdt.etal.10, Read.etal.06c, Zelnikov.Kuskov.16}. These simulation results have thus far eluded a proper explanation, and appear inconsistent with the notion that dynamical friction arises from the LBK torque which is always retarding (at least in a spherical, isotropic system, see Section~\ref{sec:torque}).

The notion that dynamical friction arises solely from resonant interactions can be traced back to the assumption that it is a secular process. This \textit{secular approximation} implies that the actions of the perturber change only very slowly, on a time scale much longer than the dynamical time. In addition, it is assumed that the perturber is introduced to the system on a long time scale (i.e., the past evolution was also secular). We shall refer to this as the \textit{adiabatic approximation}. Both the secular and adiabatic approximations underlie the treatments of TW84 and KS18, which are based on Hamiltonian perturbation, as well as all other treatments that have inferred that dynamical friction arises exclusively from resonances. This includes treatments in action-angle space that use kinetic theory and/or the fluctuation-dissipation theorem \citep[e.g.,][]{Chavanis.12, Chavanis.13, Heyvaerts.etal.17, Fouvry.Bar-Or.18}.

It is important, though, to realize that the secular and adiabatic approximations are really only justified if the dynamical friction time, defined as the timescale on which the perturber sinks to the center of its host, is much longer than the dynamical time. Such cases, though, are of limiting astrophysical interest. If instead we focus on systems for which the dynamical friction time is (significantly) shorter than the Hubble time, we are unavoidably in a regime for which the secular and adiabatic approximations may no longer be justified. In this paper we examine the impact of relaxing both these approximations. We do this by properly accounting for the past orbital evolution of the perturber in a self-consistent way. The resulting `self-consistent' torque differs from the standard LBK torque in two important ways. First of all, the self-consistent torque makes it explicitly clear that the dynamical friction torque arises from both resonant and near-resonant orbits. Secondly, while the exact resonances always exert a retarding torque, the near-resonant orbits can exert both retarding and enhancing torques. As long as the orbital decay rate is slow, the self-consistent torque can be written as the sum of two terms: (i) an `instantaneous' torque, which is the torque experienced by a perturber introduced abruptly to the host galaxy, and (ii) a `memory' torque, which depends on the entire orbital history of the perturber. The instantaneous torque builds up slowly, and then starts to oscillate in amplitude. Over time these `transient' oscillations damp out due to phase mixing, after which the instantaneous torque reduces to the LBK torque due to the pure resonant orbits. The memory torque, which always has a non-zero contribution from both resonant and near-resonant orbits, starts out sub-dominant, but becomes the dominant contributor to the total torque when the perturber approaches the core region of a galaxy. When this happens the orbital decay enters a phase of accelerated, super-Chandrasekhar infall, which ceases after the perturber crosses a critical radius at which the torque flips sign. Inside of this radius the torque is enhancing, giving rise to `dynamical buoyancy'. Hence, we argue that core stalling occurs at or near this critical radius, as a manifestation of a delicate balance between dynamical friction outside and buoyancy within. 

This paper is organized as follows. In Section~\ref{sec:standard_perturbation} we first relax the adiabatic approximation. We use Hamiltonian perturbation theory to derive an expression for the `generalized LBK torque', and discuss how it differs from the standard LBK torque using the analogy of a forced, damped oscillator. In Section~\ref{sec:general_perturbation} we subsequently also relax the secular approximation and derive an expression for the `self-consistent torque', which self-consistently accounts for the orbital evolution (decay) of the perturber. We use this torque in Section~\ref{sec:core_stalling} to discuss the orbital decay of a perturber in a cored background galaxy, providing new insight regarding core stalling, dynamical buoyancy, and super-Chandrasekhar dynamical friction. We summarize our results in Section~\ref{sec:concl}.
 
\section{Hamiltonian Perturbation Theory and the generalized LBK Torque}
\label{sec:standard_perturbation}

Throughout this paper we follow TW84 and KS18, and consider a rigid perturber\footnote{Throughout we ignore potential mass loss of the perturber due to the tidal field of the host.} of mass $M_\rmP$ moving on a circular orbit in a spherical host potential (hereafter the `galaxy') with mass profile $M_\rmG(R)$. The host is made up of a large number of `stars', or `field particles', of mass $m$, and we have that $m \ll M_\rmP \ll M_\rmG$.

\subsection{Hamiltonian Dynamics in the Co-Rotating Frame}
\label{sec:Hamilton}

Since the total, perturbed gravitational potential, and hence the Hamiltonian for each field particle, is time-variable, energy is not a conserved quantity. And due to the lack of spherical symmetry, neither is angular momentum. However, as is well known \citep[see e.g.,][]{Binney.Tremaine.08}, the Jacobi Hamiltonian
\begin{equation}
H_{\rm J} = E -{\bf \Omega_{\rm{\bP}}} \cdot \bL
\label{Ej}
\end{equation}
is a conserved quantity (if we ignore time evolution of ${\bf \Omega_{\rm{\bP}}}$). $E$ and $\bL$ are, respectively, the perturbed energy and angular momentum of the field particle in the non-rotating, inertial frame, given by
\begin{align}\label{Eperturbed}
E& =E_0+\Phi'_\rmP=\frac{1}{2}{\dot{\br}}^2 + \Phi_\rmG\left(\br\right) + \Phi'_\rmP\left(\br\right),\\
\bL &= \br \times \bf{\dot{r}}\,.
\end{align}
Here $\br$ is the position vector of the field particle with respect to the galactic center and $\dot{\br}$ is the velocity of the field particle in the inertial frame. The angular frequency of the galaxy-perturber system is given by ${\bf \Omega_{\rm{\bP}}} = (0,0,\Omega_\rmP)$, where
\begin{equation}
\Omega_\rmP = \sqrt{\frac{G\,[M_\rmG(R) + M_\rmP]}{R^3}}\,,
\end{equation}
with $R$ the galacto-centric radius of the perturber. $E_0$ is the unperturbed energy, i.e., the part of the Hamiltonian without the perturber potential, $\Phi_\rmG$ is the gravitational potential due to the galaxy, and $\Phi'_\rmP$ is the perturber potential, which consists of both direct and indirect terms and is given by
\begin{align}
\Phi'_\rmP& = \Phi_\rmP + G M_\rmP\frac{\br\cdot\bR}{R^3}\,,
\end{align}
with $\Phi_\rmP = -G M_\rmP/\vert\br-\bR\vert$ for a point perturber. The first term is the direct term, while the second term is the indirect term which accounts for the fact that the galaxy center (the origin), is rotating about the common COM with the perturber. 

In reality the perturbation in the potential also includes a gravitational `polarization' term which arises from the perturbation in the stellar distribution function induced by the perturber (also known as the `wake'). That term manifests the collective effects due to the self-gravity of the stars and significantly complicates the analysis. As first shown by \cite{Weinberg.89}, using Hamiltonian perturbation theory, and more recently by \cite{Chavanis.12}, \cite{Heyvaerts.etal.17} and \cite{Fouvry.Bar-Or.18} using the fluctuation-dissipation theorem, the collective effects primarily `dress' the perturber potential $\Phi'_\rmP$ by introducing a prefactor which is the gravitational equivalent of the dielectric function in plasma physics. In particular, as nicely summarized in \cite{Fouvry.Bar-Or.18}, taking collective effects into account in the stochastic picture yields an inhomogeneous Lenard-Balescu equation in which the diffusion coefficients involve the dressed potential\footnote{Upon using the bare potential instead (i.e., ignoring collective effects due to self-gravity of the field particles), these diffusion coefficients reduce to those of the inhomogeneous Landau equation, which in turn implies a dynamical friction force consistent with the LBK torque.}. Given the formidable challenge in computing the dressed potentials, and given that the impact of collective effects is likely less important than for plasmas \citep[see][for detailed discussion]{Chavanis.13} we follow TW84 and KS18, and neglect the effects of self-gravity for the sake of simplicity.

\bigskip

\subsection{Perturbation Analysis}
\label{sec:perturbation_analysis}

The dynamics of the unperturbed system in the co-rotating frame is governed by the Jacobi Hamiltonian $H_{0\rmJ}=E_0-{\bf \Omega_{\rm{\bP}}} \cdot \bL$, which is a conserved quantity and therefore commutes with the unperturbed distribution function $f_0$, i.e.,
\begin{align}
\left[f_0,H_{0\rmJ}\right]=0.
\end{align}
where $[A,B]$ denotes the Poisson bracket of $A$ and $B$. This is nothing but the steady state form of the collisionless Boltzmann equation (hereafter CBE) for the unperturbed galaxy in the co-rotating frame. When we introduce a perturber, the system is no longer in equilibrium and its dynamics is governed by the perturbed Hamiltonian. The perturber potential $\Phi'_\rmP$ gives rise to a perturbation in the distribution function $f_1$, which in turn exerts a torque on the perturber. This is ultimately responsible for dynamical friction\footnote{This is the key idea behind linear response theory.}. In what follows, we shall, in the spirit of KS18, perturb the CBE up to linear order to obtain the expression for $f_1$ and use it to compute the torque on the perturber.

The perturbation in the distribution function can be computed by perturbing the collisionless Boltzmann equation 
\begin{align}
\frac{\partial f}{\partial t}+[f,H_\rmJ]=0.
\label{CBE}
\end{align}
Up to linear order, we write $f$ and $H_\rmJ$ as the following perturbative series
\begin{align}
&f=f_0+f_1, \nonumber \\
&H_\rmJ=H_{0\rmJ}+\Phi_1^{\rm ext}\,. 
\label{series}
\end{align}
We follow TW84 and KS18 and let the external perturbation grow as $\Phi_1^{\rm ext}(R,t) = g(t) \, \Phi'_\rmP(R)$, where the growth function
\begin{equation}
\begin{aligned}
g(t) = 
\begin{dcases}
\rme^{\gamma t}, & t<0 \\
1,\,             & t \geq 0\,,
\label{Phi1ext}
\end{dcases}
\end{aligned}
\end{equation}
and $\gamma>0$. This indicates that the perturber grows its mass exponentially from $t\to -\infty$ to $t=0$ on a characteristic time-scale $\tau_{\rm grow} \equiv 1/\gamma$, while remaining at a fixed host-centric radius $R$. We do not consider this realistic, but before we consider an alternative we first aim to clarify the implications of this assumption. Note also that we have neglected the implicit time dependence of $\Phi_1^{\rm ext}$ and $\Omega_\rmP$ through $R(t)$. This constitutes the secular approximation that $R$ changes on a time scale much longer than the dynamical time.

Substituting the series expansions given in equation~(\ref{series}) in the CBE of equation~(\ref{CBE}), we obtain the following evolution equation for $f_1$ up to linear order
\begin{align}
&\frac{\partial f_1}{\partial t} + [f_1,H_{0\rmJ}] + [f_0,\Phi_1^{\rm ext}] = 0\,.
\label{CBEn}
\end{align}
In general, one can obtain the solution for $f_1$ once $f_0$ is known. The unperturbed distribution function $f_0$ is a solution of the unperturbed CBE and therefore, by the Jeans Theorem, is a function of the conserved quantities of the dynamical system, which in case of a spherical galaxy correspond to the three actions $I_1$, $I_2$ and $I_3$. These consist of the radial action $I_\rmr$, the total angular momentum $L$, and the $z$ component of the angular momentum, $L_\rmz$, or linear combinations thereof. Throughout, we consider the $z$-axis to coincide with the normal to the orbital plane of the perturber. In order to simplify the dynamics, we make a canonical transformation from $(\br,\bp)$ phase-space to $(\bw,\bI)$ action-angle space spanned by the action vector $\bI=\{I_1,I_2,I_3\}$ and the corresponding angle vector $\bw=\{w_1,w_2,w_3\}$. Recall that $f_0$ and $H_{0\rmJ}$ are both functions of only $I_1, I_2$ and $I_3$ and do not depend on the angles, while $f_1$ and $\Phi'_\rmP$ are functions of both actions and angles. Therefore, in action-angle space the Poisson brackets in the above equations become
\begin{align}
&[f_1,H_{0\rmJ}]=\frac{\partial f_1}{\partial w_k} \frac{\partial H_{0\rmJ}}{\partial I_k}, \\
&[f_0,\Phi'_\rmP]=-\frac{\partial f_0}{\partial I_i} \frac{\partial \Phi'_\rmP}{\partial w_i}\,.
\end{align}
Here and throughout, the Einstein summation convention is implied, and indices $k$ and $i$ run from 1 to 3 and 1 to 2, respectively. In action-angle space, $[f_1,H_{0\rmJ}]$ reduces to 
\begin{align}
[f_1,H_{0\rmJ}]=\Omega_k \frac{\partial f_1}{\partial w_k}\,,
\end{align}
where the frequencies $\Omega_k$ are given by
\begin{align}
&\Omega_1 = \frac{\partial H_{0\rmJ}}{\partial I_1} = \frac{\partial E_0}{\partial I_1}\,, \nonumber \\
&\Omega_2 = \frac{\partial H_{0\rmJ}}{\partial I_2} = \frac{\partial E_0}{\partial I_2}\,, \nonumber \\
&\Omega_3 = \frac{\partial H_{0\rmJ}}{\partial I_3} = -\Omega_\rmP\,.
\end{align}
Here $I_3=L_\rmz$, and $I_1$ and $I_2$ are linear combinations of $I_\rmr$ and $L$, respectively. Since $\Phi_\rmG$ is a spherically symmetric {potential}, $E_0$ is a function of $I_\rmr$ and $L$ only, or in other words of only $I_1$ and $I_2$. And since $f_0$ is a function of $E_0$ and $L$ only, it has a similar dependence on the actions, i.e., $\partial f_0/\partial I_3 = 0$.

Following KS18 we expand $f_1$ and $\Phi'_\rmP$ as a Fourier series in $\bw$ using
\begin{align}
&f_1(\bw,\bI,t)=\sum_{\boldell} \hat{f}_{1,\boldell}(\bI,t)\, \rme^{i\bw\cdot\boldell}, \nonumber \\
&\Phi'_\rmP(\bw,\bI)=\sum_{\boldell} \hat{\Phi}'_{\boldell}(\bI)\, \rme^{i\bw\cdot\boldell},
\label{fourier_series}
\end{align}
where the summation is over all integer triplets $\boldell = (\ell_1, \ell_2, \ell_3)$. Note that, since both $f_1$ and $\Phi'_\rmP$ are real, we have that $\hat{f}_{1,-\boldell} = \hat{f}^{\ast}_{1,\boldell}$ and $\hat{\Phi}_{-\boldell} = \hat{\Phi}^{\ast}_{\boldell}$, where $A^{\ast}$ indicates the complex conjugate of $A$. Substituting the Fourier series in equation~(\ref{CBEn}) yields the following evolution equation for $\hat{f}_{1,\boldell}$
\begin{equation}
\frac{\partial \hat{f}_{1,\boldell}}{\partial t}+i\ell_k\Omega_k \hat{f}_{1,\boldell} = g(t) \, i\ell_i \, \frac{\partial f_0}{\partial I_i} \, \hat{\Phi}'_{\boldell}\,.
\label{f1l_evolve}
\end{equation}

At this point in their analysis KS18 assume that the perturbation in the distribution function evolves in a similar way as the external perturber, i.e., $f_1 \propto \rme^{\gamma' t}$ with $\gamma' = \gamma$. Under this assumption, the above differential equation becomes a simple algebraic equation that can be solved for $\hat{f}_{1,\boldell}$. KS18 thus assume that the response density builds up on the same time scale as that on which the perturber is introduced. This is a fair assumption as long as $\gamma$ is sufficiently small, such that there is sufficient time for the host to respond. However, if dynamical friction is very efficient, then $\gamma'$ can be different from $\gamma$. In fact, in general the perturbation does not have a single growth rate. Different parts of the phase space respond to the perturber at different rates. Therefore, in what follows we will not make any a priori assumption about the growth rate of the response density due to the perturber. Rather, we solve the differential equation for $\hat{f}_{1,\boldell}$ using the Green's function technique with the initial condition that $\hat{f}_{1,\boldell}(t\to -\infty)=0$. We find the Green's function to be $e^{-i\ell_k\Omega_k \left(t-\tau\right)}$, which can be used to obtain the following particular solution for $\hat{f}_{1,\boldell}$
\begin{equation}
\begin{aligned}
\hat{f}_{1,\boldell}(\bI,t) & = i\ell_i \, \frac{\partial f_0}{\partial I_i} \, \hat{\Phi}'_{\boldell}(\bI) \, \rme^{-i\ell_k\Omega_k t}
\begin{dcases}
\int_{-\infty}^{t} \rmd\tau \, \rme^{(\gamma + i\ell_k\Omega_k)\tau}, & t<0, \\ \\
\int_{-\infty}^{0} \rmd\tau \, \rme^{(\gamma + i\ell_k\Omega_k)\tau} + \int_{0}^{t} \rmd\tau \, \rme^{i\ell_k\Omega_k\tau}, & t\geq 0.
\end{dcases}
\end{aligned}
\end{equation}
This can be integrated to yield
\begin{align}
\hat{f}_{1,\boldell}(\bI,t) & =  i\ell_i \, \frac{\partial f_0}{\partial I_i} \, \hat{\Phi}'_{\boldell}(\bI)
\begin{dcases}
\frac{ \rme^{\gamma t}}{\gamma + i\ell_k\Omega_k}, & t<0 \\ \\
\left[\frac{ \rme^{-i\ell_k\Omega_k t}}{\gamma + i\ell_k\Omega_k} + \frac{1 - \rme^{-i\ell_k\Omega_k t}}{i\ell_k\Omega_k}\right], & t \geq 0.
\end{dcases}
\label{f1l}
\end{align}
Note that the solution for $t<0$ is identical to that obtained by KS18.

\medskip

\subsection{The Generalized LBK Torque}
\label{sec:torque}

As shown in KS18, the torque on the perturber by a field particle is given by $\partial \Phi_{1,\rm ext}/\partial \phi$. Hence, the total torque on the perturber can be computed by weighting $\partial \Phi_{1, \rm ext}/\partial \phi$ by the perturbed distribution function and then integrating over all of phase space as follows
\begin{align}
\calT &= \int \rmd \br \int \rmd \bp \,\frac{\partial \Phi_1^{\rm ext}}{\partial \phi} \left(f_0+f_1\right) = \int \rmd \br \int \rmd \bp \,\frac{\partial \Phi_1^{\rm ext}}{\partial \phi} f_1.
\end{align}
Note that, since $f_0$ is independent of $\phi$ and $\oint \rmd \phi\, \left(\partial \Phi_1^{\rm ext}/\partial \phi\right)=0$, the leading order contribution to the torque comes from $f_1$. And since both $\Phi_1^{\rm ext}$ and $f_1$ are first order in $M_\rmP$, the torque itself is second-order in the mass of the perturber.

To evaluate the torque we follow KS18 and note that $\partial \Phi_1^{\rm ext}/\partial \phi = -[p_\phi,\Phi_1^{\rm ext}] = -[L_\rmz,\Phi_1^{\rm ext}]$. And since the Poisson bracket is invariant under canonical transformation, we thus have that $\partial \Phi_1^{\rm ext}/\partial \phi = -[I_3,\Phi_1^{\rm ext}] = \partial \Phi_1^{\rm ext}/\partial w_3$. Moreover the volume element $\rmd \br\, \rmd \bp$ is also invariant under canonical transformation and becomes $\rmd \bw\, \rmd \bI$. Therefore, using equation~(\ref{Phi1ext}), the torque can be written in action-angle space as
\begin{align}
\calT & = g(t) \int \rmd \bw \int \rmd \bI\, \frac{\partial \Phi'_\rmP}{\partial w_3} f_1\,,
\end{align}
After substituting the Fourier expansions of $\Phi'_\rmP$ and $f_1$, and using the reality condition, i.e. $\hat{f}_{1,-\boldell'}(\bI,t)=\hat{f}^{\ast}_{1,\boldell'}(\bI,t)$, we obtain that
\begin{align}
\calT & = g(t)  \sum_{\boldell} \sum_{\boldell'} i \ell_3 \, \int \rmd \bI\, \hat{\Phi}'_{\boldell}(\bI)  \hat{f}^{\ast}_{1,\boldell'}(\bI,t) \int \rmd \bw\,e^{i\left(\boldell-\boldell'\right)\cdot \bw},
\label{torque_intermediate}
\end{align}
which can be integrated over $\bw$ using the following identity for the Dirac delta function,
\begin{align}
\delta^3(\bx) = {1 \over {\left(2\pi\right)}^3} \int \rmd \bw\,e^{i\bw\cdot \bx},
\end{align}
and summed over the $\boldell'$ indices to yield
\begin{align}
\calT & = (2\pi)^3 \, g(t)  \sum_{\boldell} i \ell_3 \, \int \rmd \bI\, \hat{\Phi}'_{\boldell}(\bI)  \hat{f}^{\ast}_{1,\boldell}(\bI,t)\,.
\label{torque}
\end{align}
Substituting $\hat{f}_{1,\boldell}(\bI)$ given by equation~(\ref{f1l}) in the above expression, the second order torque can be written as
\begin{align}
\calT_2 = \Tgen = 16\pi^3 \reducedsum \ell_3 \int \rmd \bI\, \calJ(\ell_k\Omega_k,t)\, \ell_i\frac{\partial f_0}{\partial I_i}\, {\left|\hat{\Phi}'_{\boldell}(\bI)\right|}^2\,,
\label{Torque_gen}
\end{align}
where 
\begin{align}
\reducedsum = \sum_{\ell_1=-\infty}^{\infty} \, \sum_{\ell_2=-\infty}^{\infty} \, \sum_{\ell_3=1}^{\infty}\,, 
\end{align}
is the `reduced' summation over the positive-$\ell_3$ hemisphere of $(\ell_1,\ell_2,\ell_3)$-space, which arises from applying the mirror symmetry operation $\boldell\to -\boldell$ and retaining the symmetric part of the integrand. The function $\calJ(\ell_k\Omega_k,t)$ is given by
\begin{equation}
\begin{aligned}
\calJ(\ell_k\Omega_k, t) & = \frac{\gamma}{\gamma^2+{\left(\ell_k \Omega_k\right)}^2} 
\begin{dcases}
\rme^{2\gamma t}, & t < 0 \\ 
\cos{\ell_k\Omega_k t} + \gamma \, \frac{\sin{\ell_k\Omega_k t}}{\ell_k\Omega_k}, & t \geq 0.
\end{dcases}
\label{J}
\end{aligned}
\end{equation}
Note that this torque contains the contribution from all orbits: resonant, near-resonant and non-resonant. We therefore refer to this as the generalized LBK torque, which is based on the secular, but not the adiabatic approximation (i.e., we did not take the limit $\gamma \to 0$). The amplitude of the torque scales as the Lorentzian-like function $\gamma / [\gamma^2 + (\ell_k \Omega_k)^2]$, which peaks at the resonances, where the commensurability condition of the frequencies,
\begin{equation}
\ell_k \Omega_k = \ell_1 \Omega_1(I_1,I_2) + \ell_2 \Omega_2(I_1,I_2) - \ell_3 \Omega_\rmP = 0\,
\label{resonance_condition}
\end{equation} 
is satisfied. The width of the Lorentzian is proportional to $\gamma$, and determines the relative contributions to the torque from resonant and near-resonant orbits, with larger values of $\gamma$ (i.e., smaller $\tau_{\rm grow}$) resulting in a more dominant contribution from the near-resonant orbits.

An important feature of the generalized LBK torque is that it can be either retarding ($\calT < 0$) or enhancing ($\calT > 0$), depending on the sign of $\calJ$. Using that $\ell_i\partial f_0/\partial I_i = \ell_i\Omega_i \partial f_0/\partial E_0 < 0$ for stable distribution functions \citep[e.g.,][]{Doremus.Feix.Baumann.71, Binney.Tremaine.08}, we see that a coherent retarding (enhancing) torque corresponds to $\calJ > 0$ ($\calJ < 0$). To get some insight, we start by examining the generalized LBK torque in the limits of both small and large $\gamma$, which correspond to adiabatic growth and instantaneous introduction of the perturber, respectively.

\subsubsection{Adiabatic Growth of Perturber Potential}
\label{sec:adiabatic}

Adiabatic growth of the perturber potential implies that the perturber has to be introduced on a time scale that is long compared to all other relevant dynamical times in the problem. Only then can the distribution function $f_0(I_1,I_2,I_3)$, expressed according to the Jeans theorem as a function of its actions, remain perfectly invariant. The longest time scales of relevance are the libration times $T_{\rm lib} \sim 1/\ell_k\Omega_k$, which become infinitely long for orbits that satisfy the commensurability condition. Hence, assuring strict adiabatic invariance requires that we take the limit $\gamma \to 0$. In this limit, $\calJ(\ell_k\Omega_k,t)$ converges according to
\begin{align}
\lim_{\gamma \to 0}\calJ(\ell_k\Omega_k,t) = \pi\delta(\ell_k\Omega_k),\;\;\;\;\;-\infty<t<\infty,
\label{J_ad}
\end{align}
where we have used that the Lorentzian function becomes a Dirac delta function in the limit of vanishing width. Substituting this expression for $\calJ$ in equation~(\ref{Torque_gen}) yields the standard LBK torque
\begin{align}
\calT_{2} &= \Tlbk = 16\pi^4 \reducedsum \ell_3 \int \rmd \bI\,\delta\left(\ell_k\Omega_k\right) \left(\ell_1\Omega_1+\ell_2\Omega_2\right)\frac{\partial f_0}{\partial E_0}\, {\left|\hat{\Phi}'_{\boldell}(\bI)\right|}^2 = 16\pi^4 \Omega_\rmP \reducedsum \ell_3^2 \int \rmd \bI\,\delta\left(\ell_k\Omega_k\right)\frac{\partial f_0}{\partial E_0}\, {\left|\hat{\Phi}'_{\boldell}(\bI)\right|}^2,
\label{LBK}
\end{align}
which has a non-zero contribution from only the exact resonances. And since $\partial f_0/\partial E_0<0$ for a stable distribution function, we see that $\Tlbk < 0$ for all resonances. In other words, the standard second-order LBK torque is always retarding in nature. 

This makes it clear that the LBK torque is ultimately an outcome of taking the adiabatic limit ($\gamma \to 0$). This should not come as a surprise: in the limit where the perturber takes infinitely long to present itself, the only contribution to a net torque that is not phase mixed away (see Section~\ref{sec:analogy} and Fig.~\ref{fig:genLBK}) is that from orbits in perfect resonance with the perturber.  However, to what extent the LBK torque is relevant for dynamical friction ultimately depends on the time scale on which phase mixing removes the transient contributions. The rate at which an orbit and the perturber get out of phase depends on the incommensurability between their frequencies. If large, phase mixing is fast, and the contribution to the total torque vanishes rapidly. However, phase mixing the contribution from the near-resonant orbits can easily take many dynamical times. And since a typical galaxy or dark matter halo is only a few dynamical times old (at least in its outskirts), we are not a priori justified in assuming that dynamical friction is dominated by the LBK torque.

\subsubsection{Instantaneous Introduction of the Perturber}
\label{sec:instantaneous}

Idealized numerical simulations that examine dynamical friction  \citep[e.g.,][]{Lin.Tremaine.83, White.83, Cora.etal.97, vdBosch.etal.99, Jiang.Binney.00, Hashimoto.etal.03, Boylan-Kolchin.etal.08, Inoue.09, Inoue.11, Tamfal.etal.20} typically do not adiabatically grow the perturber potential over time, but rather introduce it instantaneously to the host system. We can use our expression for the generalized torque to examine such a scenario. Instantaneous introduction of the perturber corresponds to taking $\gamma \to \infty$, for which 
\begin{align}
\lim_{\gamma \to \infty}\calJ(\ell_k\Omega_k,t) = 
\begin{dcases}
0, & t<0, \nonumber \\ \\
\frac{\sin{\ell_k\Omega_k t}}{\ell_k\Omega_k}, & t\geq 0.
\end{dcases}
%\label{J_ad}
\end{align}
This yields the following expression for the torque 
\begin{align}
\calT_{2} &= \Tinst \equiv 16\pi^3 \reducedsum \ell_3 \int \rmd \bI\,\frac{\sin{\ell_k\Omega_k t}}{\ell_k\Omega_k} \left(\ell_1\Omega_1+\ell_2\Omega_2\right)\frac{\partial f_0}{\partial E_0}\, {\left|\hat{\Phi}'_{\boldell}(\bI)\right|}^2.
\label{Torque_inst}
\end{align}
We shall hereafter refer to this as the instantaneous torque. At small $t$, i.e., shortly after the instantaneous introduction of the perturber, we have that $\Tinst \propto t$, indicating that the second-order torque builds up linearly with time. Note that the contribution to the torque from the non-resonance orbits can be either retarding or enhancing. In particular, for modes that are sufficiently far away from resonance, $\sin{\ell_k\Omega_k t}$ can become negative a short time after the instantaneous introduction of the perturber, which can result in an enhancing torque. In addition, since $\ell_1 \Omega_1 + \ell_2 \Omega_2$ can be either positive or negative, it is even possible (at least in principle) for $\Tinst$ to be enhancing when $\sin{\ell_k\Omega_k t}$ is positive.

Finally, note that in the $t \to \infty$ limit, $\sin{(\ell_k\Omega_k  t)} / \ell_k\Omega_k \to \pi \, \delta(\ell_k\Omega_k)$, and we recover the familiar LBK formula for the torque, with a non-zero contribution only from the exact resonances. Hence, in accordance with the analogy of the forced, damped oscillator discussed in Section~\ref{sec:analogy}, following the instantaneous introduction of the perturber, the torque initially builds up linearly with time, then undergoes oscillations (corresponding to transients) that slowly phase mix away, after which only the LBK torque due to the perfect resonances remains. 

\begin{figure*}
\centering
\includegraphics[width=0.8\textwidth]{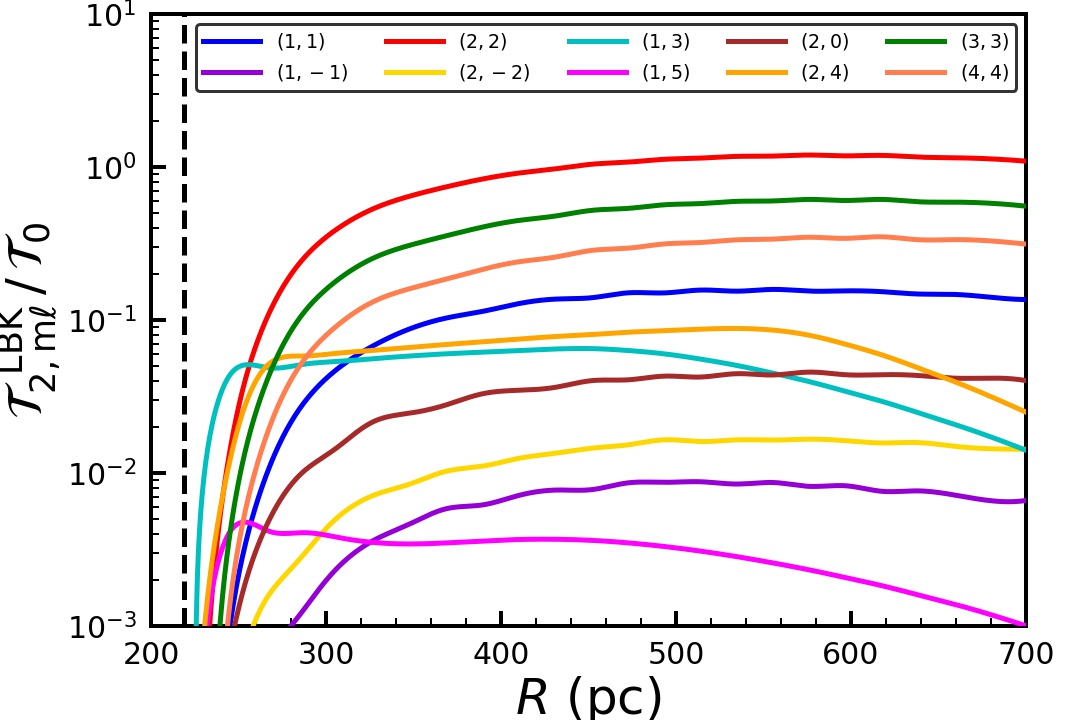}
\caption{The LBK torque on a point mass perturber of mass $M_\rmP$ on a circular orbit in a spherical isochrone potential of mass $M_\rmG = 8000 M_\rmP$, in units of $\calT_0 \equiv G M^2_\rmP/b$, as a function of the galacto-centric radius of the perturber, $R$. Different curves show the contribution due to the ten $(m,l) = (m,l,m)$ resonance orbits (modes) that dominate the total LBK torque, as indicated. Note how all the $m=|l|$ modes contribute a torque with a similar $R$-dependence, and that the LBK torque dies out as the perturber approaches the `filtering radius' $R_\ast= 0.22\,b = 220\pc$, indicated by the black vertical, dashed line. As discussed in KS18, this decline of the (LBK) torque as the perturber approaches the central core region is responsible for the phenomenon of core stalling (but see section~(\ref{sec:core_stalling}) for a somewhat different explanation).}
\label{fig:LBKiso}
\end{figure*}

\subsection{Dynamical Friction in an Isochrone Sphere}
\label{sec:iso}

In order to illustrate how the instantaneous torque differs from the LBK torque, we compare the two for the case of a point mass perturber on a circular orbit in a spherical, isotropic isochrone galaxy \citep[][]{Henon.59}. This configuration was also considered by KS18 and has the advantage that (i) it is a fairly realistic representation of a galaxy, (ii) many of its physical quantities can be computed analytically, and (iii) it has a central constant density core, which allows us to examine core stalling. 

The gravitational potential of Henon's isochrone sphere with mass $M_\rmG$ and scale radius $b$ is given by
\begin{align}\label{IsoPot}
\Phi_\rmG = -\frac{GM_\rmG}{b+\sqrt{b^2+r^2}}\,,
\end{align}
and its corresponding density profile, 
\begin{align}\label{IsoDens}
\rho_\rmG(r) &= \frac{M_\rmG}{4\pi} \left[\frac{3\left(b^2+r^2\right)\left(b+\sqrt{b^2+r^2}\right)-r^2\left(b+3\sqrt{b^2+r^2}\right)}{{\left(b+\sqrt{b^2+r^2}\right)}^3{\left(b^2+r^2\right)}^{3/2}}\right]
\end{align}
falls off as $r^{-4}$ at large $r$, and asymptotes to a constant core value of $3M_\rmG/16\pi b^3$ as $r \rightarrow 0$. Following KS18, we adopt $M_\rmG = 1.6\times 10^9 \Msun$ and $b=1\kpc$. These parameters were chosen by KS18 such that the isochrone sphere has the same core radius and central density as the \citet{Burkert.95} sphere used in the high-resolution $N$-body simulation of \citet{Inoue.11}. Following both  KS18 and \citet{Inoue.11}, we adopt a point mass perturber of mass $M_\rmP = 2\times 10^5 \Msun$ (corresponding to a mass ratio $M_\rmP/M_\rmG = 1.25\times 10^{-4}$), which we consider to be on a circular orbit. In what follows we shall refer to this set-up as our fiducial example.

As detailed in Appendix~\ref{app:model} and KS18, the commensurability condition for this system can be written as
\begin{align}
\ell_k\Omega_k = n\,\Omega_w + l\,\Omega_g - m\,\Omega_\rmP 
%\ell_k\Omega_k = m\,\Omega_w + \ell\,\Omega_g - n\,\Omega_\rmP 
\end{align}
where, following KS18, we have used $\ell_1 = n$, $\ell_2=l$ and $\ell_3 =m$. The frequencies $\Omega_w$ and $\Omega_g$ are related to the radial and angular frequencies in the orbital plane, as described in Appendix~\ref{app:model}. Although the total (generalized) LBK torque is the sum over all $(n,l,m)$, KS18 have shown that the torque is dominated by the co-rotation resonances, which have $m=n$. In addition, the torque is typically stronger for lower order modes $(|l| \lsim 3m)$. In what follows we therefore restrict ourselves to the $(m,l) = (m,l,m)$ modes with dominant LBK torque. 

Fig.~\ref{fig:LBKiso} plots the LBK torque (computed using equation~[\ref{Torque_LBK_isochrone}] as detailed in  Appendix~\ref{app:model}), as a function of the galacto-centric distance of the perturber, $R$, for the 10 dominant $(m,l)$ modes. Note how the LBK torque is dominated by that due to the $(m,l) = (2,2)$ resonance, and that the LBK torque dies out as the perturber approaches the `filtering radius' $R_\ast = 0.22\, b = 220 \pc$, marked by the black vertical dashed line. As detailed in KS18, at this radius, the circular frequency $\Omega_\rmP$ equals $\Omega_b = 0.5\sqrt{G M_\rmG/b^3}$, roughly the circular frequency of stars in a central core of the isochrone sphere. As a result, the phase space contributing to the resonances shrinks, strongly suppressing the contribution of the dominant, lower order modes to the total torque.

\begin{figure*}
\centering
\includegraphics[width=1\textwidth]{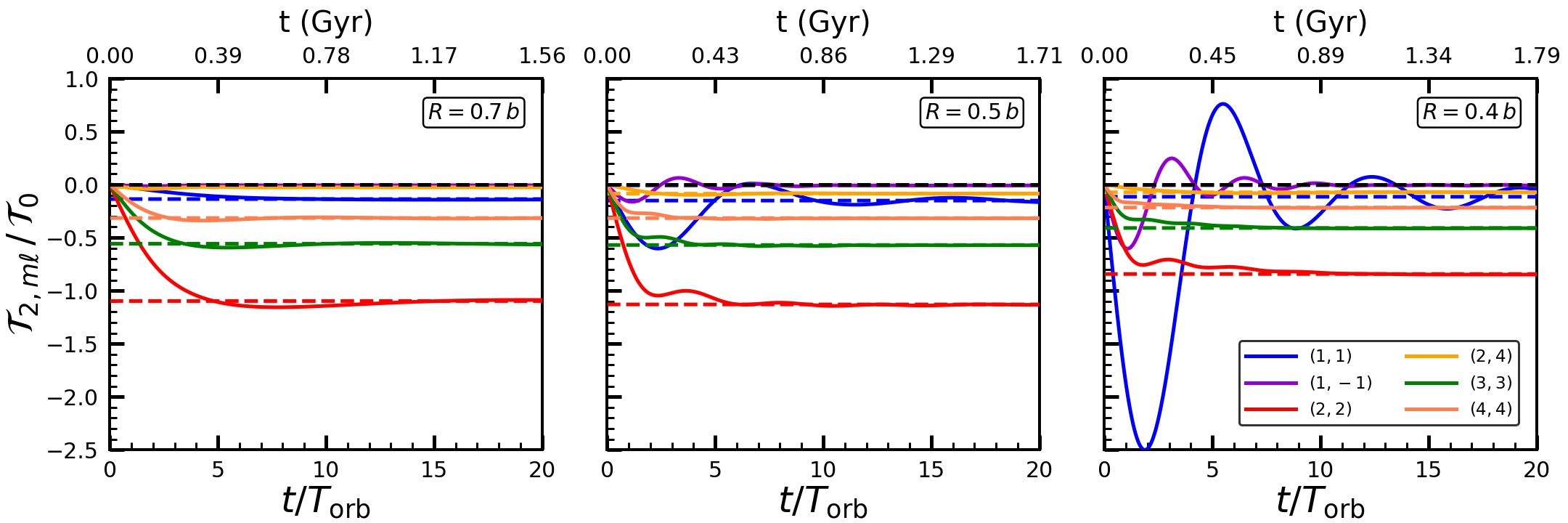}
\caption{The instantaneous, generalized LBK torque (assuming an isochrone model for the galaxy and point perturber with $M_\rmP/M_\rmG=1.25\times 10^{-4}$), in units of $\calT_0=G M^2_\rmP/b$ as a function of $t/T_{\rm orb}$ ($T_{\rm orb}=2\pi/\Omega_\rmP$ is the orbital period of the perturber) when the perturber is introduced at $R=0.7\, b$ (left panel), $0.5\, b$ (middle panel) and $0.4\,b$ (right panel). The solid lines show the instantaneous torque for six of the dominant modes as indicated. The dashed lines show the corresponding LBK torque. Note that the instantaneous torque converges to the LBK torque as $t\to\infty$ as all but the perfectly resonant orbits get phase-mixed away.}
\label{fig:genLBK}
\end{figure*}

The solid lines in Fig.~\ref{fig:genLBK} plot the instantaneous torque of equation~(\ref{Torque_inst_isochrone}) as a function of time in units of $T_{\rm orb} = 2\pi/\Omega_\rmP$ for six modes that dominate the total torque either at early and/or late times. Results are shown for three different radii of introduction of the perturber, $R=0.7\, b$ (left panel), $0.5\, b$ (middle panel) and $0.4\, b$ (right panel). For comparison, we also plot the corresponding LBK torque as horizontal dashed lines. All modes initially show a coherent, retarding torque, causing the torque to build-up linearly with time, before undergoing oscillations about the corresponding LBK value that slowly damp away with time. This is a classic example of phase-mixing in collisionless systems where all but the purely resonant responses in the distribution function are damped out. Note, though, that the transients from some modes take many orbital times to phase-mix away, especially at smaller radii (closer to the core). The reason is that the core region with a nearly constant density has a much narrower dynamic range in orbital frequencies, resulting in a more coherent response. Interestingly, the transient oscillations can even contribute a positive, enhancing torque at times. Therefore, while the late time dynamics is governed by the LBK torque from the perfectly resonant orbits, the transient behaviour following the introduction of the perturber is driven by the near-resonant orbits. 

\bigskip

\subsection{Analogy: the Forced, Damped Oscillator}
\label{sec:analogy}

It is insightful to compare the galaxy-plus-perturber system to a forced, damped oscillator. As is well known, the general solution of a sinusoidally forced, damped oscillator is a sum of a transient solution that depends on the initial conditions and a steady state solution that is independent of the initial conditions. With time, the transients damp away, causing the response to settle towards the steady state solution, which is a sinusoidal oscillation with a frequency equal to that of the driver, and an amplitude that depends on the driving amplitude, the driving frequency, the eigenfrequency of the (undamped) oscillator, and the damping ratio.

This is similar to how a galaxy responds to a perturber, $M_\rmP$, on a (circular) orbit with frequency $\Omega_\rmP$, which can be regarded as the forcing frequency. The galaxy, in turn, acts as a damped oscillator. In fact, the galaxy is an ensemble of many individual oscillators (the individual orbits), each with three frequencies (corresponding to the three actions). Although each of these orbital frequencies can be considered to correspond to an individual {\it undamped} oscillator, the collective response of all orbits acts as if damped. The source of damping is phase-mixing; initially, when the perturber is introduced (i.e., the forcing commences), all near-resonant orbits respond in phase, and the galaxy response is dominated by transient behavior. Due to phase-mixing, though, the responses of different orbits get out of phase, causing the transient behavior to die out. The only orbits that never get out of phase with the forcing are the resonant orbits, which due to the commensurability of their orbital frequencies with $\Omega_\rmP$, remain in phase, thereby resisting phase mixing. As a consequence, the `steady-state response' of the galaxy is the LBK torque due to the resonance orbits.

This analogy shows that taking the limit $\gamma \to 0$ corresponds to `skipping' the transient behavior, assuming that phase mixing has caused the response of all non-resonant orbits to die out. However, it is clear from the analogy that one can ignore the transients only after a sufficiently long time. If the perturber inspirals rapidly (i.e., $\dot{\Omega}_\rmP$ is large), then the forcing with frequency $\Omega_\rmP$ may not last sufficiently long for phase mixing to nullify the net response of all non-resonant orbits. In particular, the near-resonant orbits, whose response takes the longest to phase mix away, are expected to make a significant, if not dominant, contribution. The generalized LBK torque includes the transient response due to non-resonant orbits and presents a proper description of how dynamical friction builds up in idealized numerical simulations that introduce the perturber instantaneously. However, it is obtained by only relaxing the adiabatic approximation while still relying on the secular approximation, i.e. the perturber undergoes a slow inspiral under dynamical friction. In the following section we relax this assumption and develop a fully self-consistent treatment for dynamical friction, that includes the `memory effect' due to the entire past orbital history of the perturber.

\bigskip

\section{Self-consistent computation of the torque}
\label{sec:general_perturbation}

The previous section has given some useful insight as to how transients that result from an instantaneous introduction of the perturber ($\gamma \to\infty$) phase mix away, ultimately giving rise to the LBK torque that one obtains in the adiabatic limit ($\gamma \to 0$). However, that entire analysis is based on the generalized LBK torque (equations~[\ref{Torque_gen}]--[\ref{J}]), which assumes that the perturber grows its mass exponentially, on a time scale  $\tau_{\rm grow} = 1/\gamma$, while continuing to orbit at a fixed radius $R$. This is not realistic. The proper way to introduce the perturber is to self-consistently account for its past trajectory $R(t)$, from $t=0$ to the present, which is what we tackle in this section.

The main shortcoming with the derivation of the generalized LBK torque,
or with that of the standard LBK torque in TW84 and KS18, is that it ignores the time dependence of $\Phi'_\rmP$ and $\Omega_\rmP$ in solving the evolution equation for $\hat{f}_{1,\boldell}$ (equation~[\ref{f1l_evolve}]). In this section we are going to relax this assumption of secular evolution and compute $\hat{f}_{1,\boldell}$ and ultimately the torque in a fully self-consistent way. Here we assume that the perturber starts out at $t=0$ at some large initial radius, $R_0$, and slowly makes its way inwards following a circular orbit with time-dependent radius $R(t)$. Therefore, both the external perturbation and its circular frequency depend implicitly on time according to
\begin{align}
\Phi_1^{\rm ext} = \Phi'_\rmP(R(t)),\;\;\;\;\;\;\;\;\;\;
\Omega_\rmP = \Omega_\rmP (R(t))\,.
\end{align}
In what follows, for brevity, we simply write these dependencies as $\Phi'_\rmP(t)$ and $\Omega_\rmP(t)$ and consider it understood that the time-dependence enters implicitly via the perturber's orbit, $R(t)$.

As in \S\ref{sec:standard_perturbation}, we can perturb the CBE up to linear order and expand $f_1$ and $\Phi'_\rmP$ as the Fourier series of equation~(\ref{fourier_series}) to obtain the following evolution equation for $\hat{f}_{1,\boldell}$
\begin{align}
\frac{\partial \hat{f}_{1,\boldell}}{\partial t} + i\left[\ell_i\Omega_i - \ell_3\Omega_\rmP(t)\right] \hat{f}_{1,\boldell} = i\ell_i\frac{\partial f_0}{\partial I_i}\hat{\Phi}'_{\boldell}(\bI,t),
\end{align}
where, as before, the index $i$ runs from 1 to 2. The above equation can be solved using the Green's function with the initial condition that $\hat{f}_{1,\boldell}(\bI,0)=0$ to yield the following form for $\hat{f}_{1,\boldell}(\bI,t)$
\begin{align}
\hat{f}_{1,\boldell}(\bI,t) = i\ell_i\frac{\partial f_0}{\partial I_i}\int_{0}^t \rmd \tau e^{-i\zeta(\tau)} \, \hat{\Phi}'_{\boldell}(\bI,t-\tau),
\end{align}
where
\begin{align}
\zeta(\tau) = \ell_i \Omega_i \tau - \ell_3 \int_0^{\tau} \Omega_\rmP(t') \rmd t'.
\end{align}
We substitute this expression in equation~(\ref{torque}) without the $g(t)$ factor to obtain the following form for the self-consistent torque
\begin{equation}
\calT_{2} = \Tsc = 16\pi^3 \reducedsum \ell_3 \int \rmd \bI\, \ell_i\frac{\partial f_0}{\partial I_i}\, \left[\calJ_{1\boldell}(\bI,t)+\calJ_{2\boldell}(\bI,t)\right]\,.
\label{Torque_SC}
\end{equation}
Here $\calJ_{1\boldell}(\bI,t)$ and $\calJ_{2\boldell}(\bI,t)$ are given by
\begin{align}
\calJ_{1\boldell}(\bI,t) = \rm{Re}\left[ \hat{\Phi}^{'\ast}_{\boldell}(\bI,t) \int_{0}^t \rmd \tau \cos\zeta(\tau) \, \hat{\Phi}'_{\boldell}(\bI,t-\tau)\right],\nonumber \\
\calJ_{2\boldell}(\bI,t) = \rm{Im}\left[ \hat{\Phi}^{'\ast}_{\boldell}(\bI,t) \int_{0}^t \rmd \tau \sin\zeta(\tau) \, \hat{\Phi}'_{\boldell}(\bI,t-\tau)\right],
\label{calJsc}
\end{align}
where $\rm{Re}(z)$ and $\rm{Im}(z)$ are the real and imaginary parts of $z$, respectively. As we shall see shortly, $\calJ_{1,\boldell}$ is generally the dominant term and $\calJ_{2,\boldell}$ is sub-dominant. 

Equation~(\ref{Torque_SC}) is the most general form for the dynamical friction torque in the framework of linear perturbation theory in the absence of collective effects. This self-consistent torque differs from the instantaneous, generalized LBK torque of equation~(\ref{Torque_inst}) in two important ways. First of all, it  modifies the resonances by introducing a time-dependence to the circular frequency $\Omega_\rmP$. Mathematically, this implies that the argument $\ell_k\Omega_k \tau$ of the sinusoidal function in the instantaneous torque is replaced by $\ell_i \Omega_i \tau - \ell_3 \int_0^{\tau} \Omega_\rmP(t') \rmd t'$. Secondly, the self-consistent torque properly accounts for the fact that the perturber potential $\Phi'_\rmP$ evolves as the perturber falls in. This implies that the ${|\hat{\Phi}'_{\boldell}(R(t))|}^2$ term in the instantaneous torque is replaced by a convolution term, i.e.
\begin{align}
\frac{\sin{\ell_k\Omega_k t}}{\ell_k\Omega_k} {\left|\hat{\Phi}'_{\boldell}(\bI,t)\right|}^2 \to \rm{Re} \left[\hat{\Phi}^{'\ast}_{\boldell}(\bI,t)\int_{0}^t \rmd \tau  \cos\zeta(\tau) \, \hat{\Phi}'_{\boldell}(\bI,t-\tau)\right].
\end{align}
In the secular limit where the temporal evolution is very slow, such that $\hat{\Phi}'_{\boldell}(t-\tau) \approx \hat{\Phi}'_{\boldell}(t)$ and $\Omega_\rmP(t') \approx \Omega_\rmP(t)$, we have that
\begin{align}
\calJ_{1,\boldell}(\bI,t) \approx  {\left|\hat{\Phi}'_{\boldell}(\bI,t)\right|}^2 \int_0^t \rmd\tau \cos{\ell_k\Omega_k \tau} = {\left|\hat{\Phi}'_{\boldell}(\bI,t)\right|}^2 \, \frac{\sin{\ell_k\Omega_k t}}{\ell_k\Omega_k},
\end{align}
while $\calJ_{2,\boldell}\approx 0$. Substituting this in equation~(\ref{Torque_SC}), one recovers the expression for the instantaneous torque of equation~(\ref{Torque_inst}), as required.

While the instantaneous torque only depends on the current time $t$, the self-consistent torque takes into account the entire infall history of the perturber, thereby introducing temporal correlation into the system. This is reminiscent of how in the linear response theory of \cite{Colpi.Pallavicini.98} the overdensity along the trail marked by the perturber exerts a retarding torque on it, i.e., dynamical friction originates from a memory effect involving the stars along the path of the perturber. The self-consistent torque properly accounts for this memory effect, which is ignored in both the instantaneous torque and the LBK torque. While \cite{Colpi.Pallavicini.98} compute the torque for an impulsive, straight orbit of the perturber through a homogeneous medium, our self-consistent torque (equation~[\ref{Torque_SC}]) describes dynamical friction for the more realistic case of a circular orbit in an inhomogeneous background. We also emphasize that even the inhomogeneous Lenard-Balescu equation derived by \cite{Heyvaerts.10}, \cite{Chavanis.12} and \cite{Fouvry.Bar-Or.18}, which is considered the most complete kinetic theory for gravitational systems to date, accounting for both inhomogeneity and collective effects, ignores the memory effect by assuming, in the computation of the diffusion coefficients, that the motion of the perturber is given by the mean-field limit with time-invariant actions (the secular approximation).

\bigskip

\subsection{Orbital Decay}
\label{sec:orbdec}

Under the assumption that the evolution of $R(t)$ is governed by the second-order dynamical friction torque, we have that
\begin{align}
\frac{\rmd R}{\rmd t} = \frac{\rmd R}{\rmd L_\rmP} \, \frac{\rmd L_\rmP}{\rmd t} = \left(M_\rmP\frac{\rmd}{\rmd R}\left[R^2\Omega_\rmP(R)\right]\right)^{-1} \calT_2,
\label{dRdt_torque}
\end{align}
where $L_\rmP=M_\rmP R^2\Omega_\rmP$ is the angular momentum of the perturber. And since the torque $\calT_2$ itself depends on time both explicitly and implicitly (through $R(t)$), we have that the evolution of $R$ is governed by the following integro-differential equation
\begin{align}
&\frac{\rmd R}{\rmd t} = 16\pi^3 \left(M_\rmP\frac{\rmd}{\rmd R}\left[R^2\Omega_\rmP(R)\right]\right)^{-1} \reducedsum \ell_3 \int \rmd \bI\, \ell_i\frac{\partial f_0}{\partial I_i}\, \left[\calJ_{1\boldell}(\bI,t)+\calJ_{2\boldell}(\bI,t)\right].
\label{dRdt}
\end{align}

Solving this equation is rather challenging. However, we can obtain some powerful insight by expanding $R(t)$ using a Taylor series expansion. As long as the rate of infall,  $\rmd R/\rmd t$, varies sufficiently slowly (i.e., $\rmd^2 R/\rmd t^2$ is small), we have that, to good approximation, $R(t-\tau) \approx R(t) - \tau\,\rmd R/\rmd t$. Since $\rmd R/\rmd t \sim M_\rmP/M_\rmG$ and is therefore typically small, $\hat{\Phi}'_{\boldell}(t-\tau)$ can be expanded as a Taylor series and truncated at the leading order to obtain
\begin{align}
\hat{\Phi}'_{\boldell}(t-\tau) \approx \hat{\Phi}'_{\boldell}\big( R - \tau\,\rmd R/\rmd t\big) \approx \hat{\Phi}'_{\boldell}(t) - \frac{\rmd \hat{\Phi}'_{\boldell}}{\rmd R} \, \frac{\rmd R}{\rmd t} \, \tau\,,
\end{align}
where we remind the reader that the time dependence of $\hat{\Phi}'_{\boldell}$ only enters through $R(t)$. Next we note that $\zeta(\tau) = \tau\left[\ell_1 \Omega_1 + \ell_2 \Omega_2 - \ell_3 \overline{\Omega}_\rmP(\tau)\right]$, where
\begin{align}
\overline{\Omega}_\rmP(\tau) = \frac{1}{\tau} \int_0^{\tau} \Omega_\rmP(t') \rmd t'
\label{barOmegaP}
\end{align}
is the time-averaged value of the circular frequency of the perturber. If we now make the assumption that $\overline{\Omega}_\rmP \simeq \Omega_\rmP$, i.e., we neglect the temporal evolution of $\Omega_\rmP$\footnote{This is a reasonable approximation for a low-mass perturber on a circular orbit inside or close to a constant density core, which is\\ the case of interest here.}, then $\zeta(\tau) \to \ell_k\Omega_k\tau$. We thus have that $\calJ_{2,\boldell}\approx 0$, and 
\begin{align}
\calJ_{1,\boldell}(\bI,t) &\approx  {\left|\hat{\Phi}'_{\boldell}(\bI,t)\right|}^2 \int_0^t \rmd\tau \cos{\ell_k\Omega_k \tau} - \frac{1}{2} \frac{\rmd \vert\hat{\Phi}'_{\boldell}(\bI,t)\vert^2}{\rmd R}\, \frac{\rmd R}{\rmd t} \int_0^t \rmd\tau\,\tau\cos{\ell_k\Omega_k \tau} \nonumber \\
\nonumber \\
&= {\left|\hat{\Phi}'_{\boldell}(\bI,t)\right|}^2\,\frac{\sin{\ell_k\Omega_k t}}{\ell_k\Omega_k} - \frac{1}{2} \frac{\rmd \vert\hat{\Phi}'_{\boldell}(\bI,t)\vert^2}{\rmd R} \, \frac{\rmd R}{\rmd t}\left(t\frac{\sin{\ell_k\Omega_k t}}{\ell_k\Omega_k}-\frac{1-\cos{\ell_k\Omega_k t}}{{\left(\ell_k\Omega_k\right)}^2}\right).
\label{J1l}
\end{align}
Substituting the above expression for $\calJ_{1,\boldell}$ in equation~(\ref{Torque_SC}), we can write the second-order self-consistent torque as
\begin{equation}
\calT_{2} = \Tinst + \Tmem,
\label{Torque_SC_fast}
\end{equation}
where
\begin{equation}
\Tmem = -8 \pi^3 \frac{\rmd R}{\rmd t} \reducedsum \ell_3 \int \rmd \bI\, \ell_i \frac{\partial f_0}{\partial I_i}\, \frac{\rmd \vert\hat{\Phi}'_{\boldell}(\bI,t)\vert^2}{\rmd R} \,\left(t\frac{\sin{\ell_k\Omega_k t}}{\ell_k\Omega_k} - \frac{1-\cos{\ell_k\Omega_k t}}{{\left(\ell_k\Omega_k\right)}^2}\right)\,.
\label{Torque_mem}
\end{equation}
Hence, the torque is the sum of the instantaneous torque given by equation~(\ref{Torque_inst}), and a leading order correction term due to the inward radial motion of the perturber. In what follows, we refer to this second term as the memory term.

We can substitute the above expression for the torque (equation~[\ref{Torque_SC_fast}]) in equation~(\ref{dRdt_torque}) to obtain the following evolution equation for $R$,
\begin{align}
\dfrac{\rmd R}{\rmd t} = \dfrac{\Tinst}{\rmd L_\rmP/\rmd R - \Pmem}
\label{dRdt_selfconst}
\end{align}
where $\Pmem \equiv \Tmem/(\rmd R/\rmd t)$ is a momentum term associated with the orbital decay (i.e., the `sinking') of the perturber, and 
\begin{align}
\frac{\rmd L_\rmP}{\rmd R} = \frac{L_\rmP}{R} \, \left[2 + \frac{\rmd\ln\Omega_\rmP}{\rmd\ln R} \right]
\end{align}
is related to the momentum of the perturber in the absence of orbital decay.

At small $t$, $\Pmem$ is small compared to $\rmd L_\rmP/\rmd R$ and the infall is driven by the instantaneous torque, which is subject to transients that slowly die out due to phase mixing. As time goes on, and orbital decay becomes significant, $\Pmem$ which we find to be typically positive, becomes more and more important, causing the denominator to become smaller. This in turn enhances the orbital decay rate. Hence, the memory term of the self-consistent torque has a destabilizing effect on the orbital decay. This is similar to the destabilizing `dynamical feedback' discussed in TW84. As we will see below, this becomes particularly important when the perturber approaches a central constant density core.

\begin{figure*}
\centering
\includegraphics[width=0.6\textwidth]{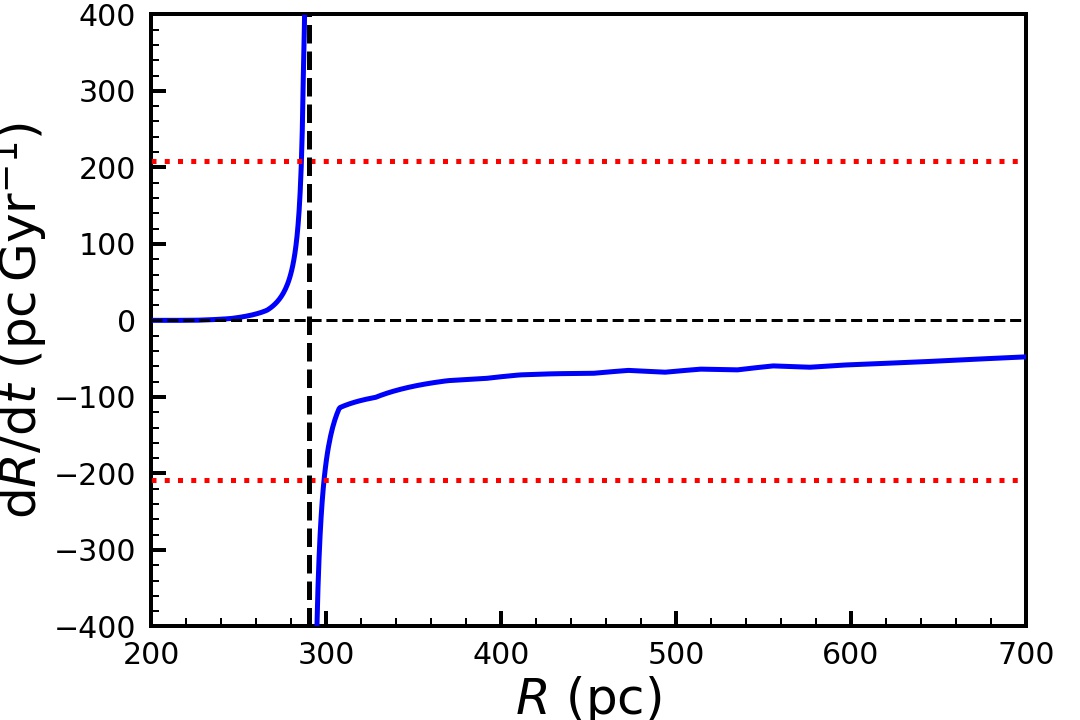}
\caption{The orbital decay rate, $\rmd R/\rmd t$, for our fiducial isochrone plus point-mass perturber system as a function of radius $R$ (equation~\ref{dRdt_selfconst}) for the asymptotic (large $t$) value of  the self-consistent torque exerted by the 10 dominant $(m,l)$ modes shown in Fig.~\ref{fig:LBKiso}. Note that under the approximation of a linear order truncation in $\calJ_{1,\boldell}$ as assumed in deriving equation~(\ref{dRdt_selfconst}), $\rmd R/\rmd t \to \pm \infty$ as $R\to R_{\rm crit}=290$ pc (marked by the black vertical dashed line) from left or right. In order to avoid this singular behavior when calculating the orbital decay, we implement a maximum cut-off for $\left|\rmd R/\rmd t\right|$, indicated by the red, dotted lines.}
\label{fig:dRdt_vs_R}
\end{figure*}

Note that equation~(\ref{dRdt_selfconst}) indicates the potential presence of a singularity at a critical radius, $\Rcrit$, where $\rmd L_\rmP/\rmd R = \Pmem$. Whether such a radius exists or not depends on both the galaxy potential and the mass of the perturber\footnote{In absence of the perturber, for a cored profile, $\rmd L_\rmP/\rmd R \sim R$ for small $R$. $\Pmem$ on the other hand, typically increases with\\ decreasing $R$. Therefore, for small enough $R$, $\Pmem$ will overtake $\rmd L_\rmP/\rmd R$ and $\rmd R/\rmd t$ will flip sign.}. In the limit $R \downarrow \Rcrit$, the orbital decay rate $\rmd R/\rmd t \to -\infty$. Inside of $\Rcrit$, the denominator flips sign and $\rmd R/\rmd t \to +\infty$ as $R \uparrow \Rcrit$. Fig.~\ref{fig:dRdt_vs_R} demonstrates this by plotting $\rmd R/\rmd t$ as a function of radius $R$ for our fiducial isochrone galaxy plus point mass perturber (see Section~\ref{sec:iso} for details). Here we have assumed the asymptotic (large time) forms for both the instantaneous torque (which equates to the LBK torque) and for the `memory torque', $\Tmem$. The fact that $\rmd R/\rmd t$ flips sign when crossing $\Rcrit$ suggests that this radius must act as an attractor for the dynamical evolution of the perturber, and we therefore associate $\Rcrit$ with the `core-stalling' radius.

\bigskip

\section{Super-Chandrasekhar dynamical friction, Dynamical buoyancy and core-stalling}
\label{sec:core_stalling}

For our fiducial example of a point mass perturber of mass $M_\rmP = 2\times 10^5 \Msun$ on a circular orbit in a spherical isochrone galaxy of mass  $M_\rmG = 1.6\times 10^9 \Msun$ and scale radius $b=1\kpc$ (see Section~\ref{sec:iso}), we use equation~(\ref{dRdt_selfconst}) and a fourth order Runge-Kutta integrator to evolve the radius $R(t)$ of the perturber.  As before, we only consider the contribution to the torque from the ten dominant $(m,l)$ modes shown in Fig.~\ref{fig:LBKiso}. We have verified that this sampling of only the dominant modes does not significantly impact the results; in fact, we obtain virtually identical results if we were to only use the eight most dominant modes. In order to avoid problems with the integrator close to the singularity at $\Rcrit \simeq 0.29 b = 290\pc$, we implement a maximum cut-off in $\vert \rmd R/\rmd t\vert$ (shown by dotted, red lines in Fig.~\ref{fig:dRdt_vs_R}).

\begin{figure*}
\centering
\includegraphics[width=0.9\textwidth]{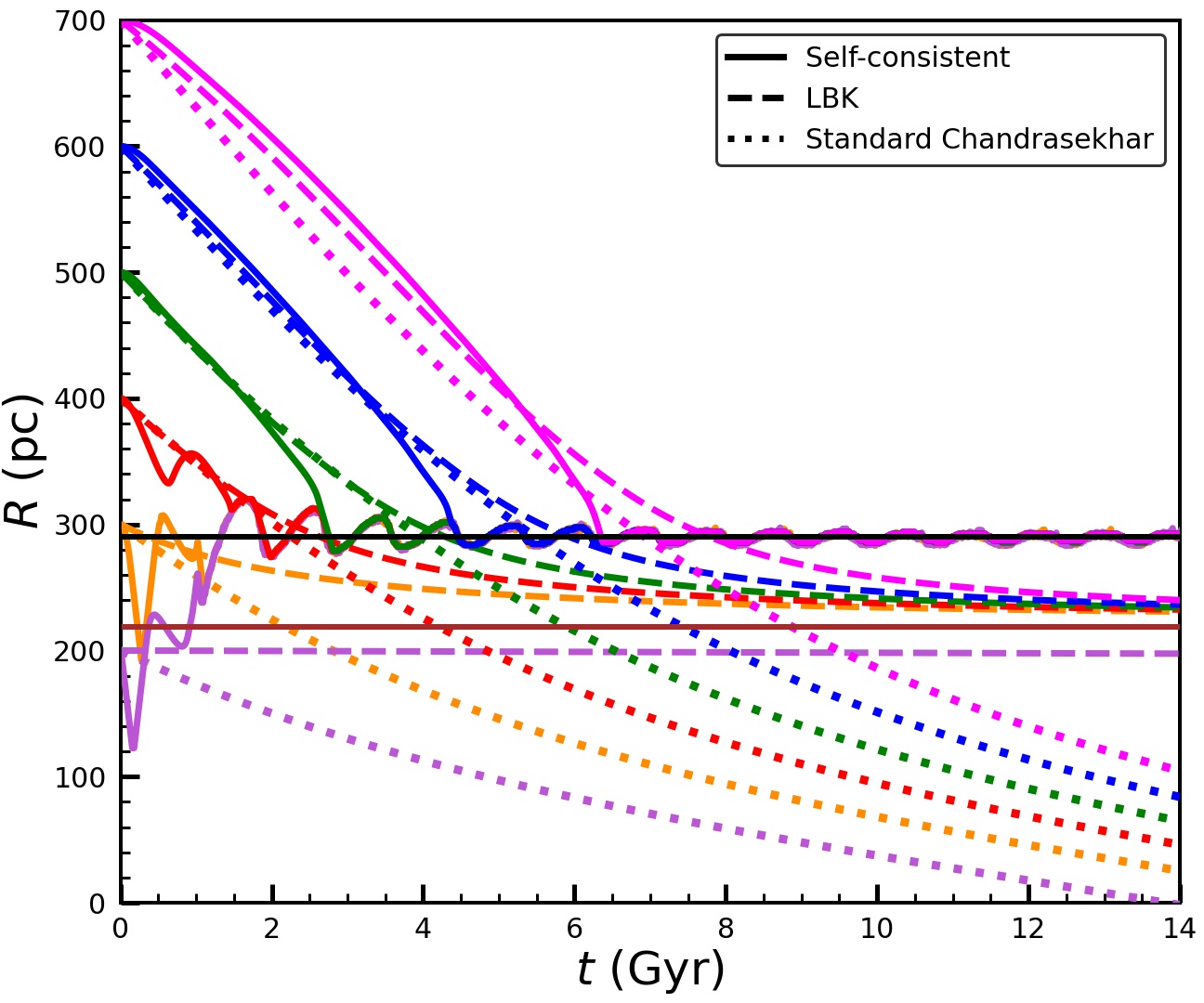}
\caption{The orbital decay of a point mass perturber in our fiducial isochrone sphere. Solid and dashed lines show the results obtained using the self-consistent and LBK torques, respectively, computed using the 10 dominant $(m,l)$ modes shown in Fig.~\ref{fig:LBKiso}. The dotted curves show the results obtained using the standard Chandrasekhar formalism, as described in the text. Different colors correspond to different initial radii $R_0 = 700\pc, 600\pc,..., 200\pc$. The horizontal black line indicates the critical radius, $R_{\rm crit}$, where the perturber stalls its infall in our self-consistent formalism. Note the transients at early times when $R_0 \sim R_{\rm crit}$, and the super-Chandrasekhar decay shortly before stalling. For comparison, based on the LBK torque stalling happens at the somewhat smaller filtering radius, $R_\ast$ (horizontal, brown line), defined in KS18 as the radius where $\Omega_\rmP(R)=\Omega_b$. Note that no stalling is expected with the standard Chandrasekhar formalism.}
\label{fig:R_vs_t}
\end{figure*}

\begin{figure*}
\centering
\includegraphics[width=1\textwidth]{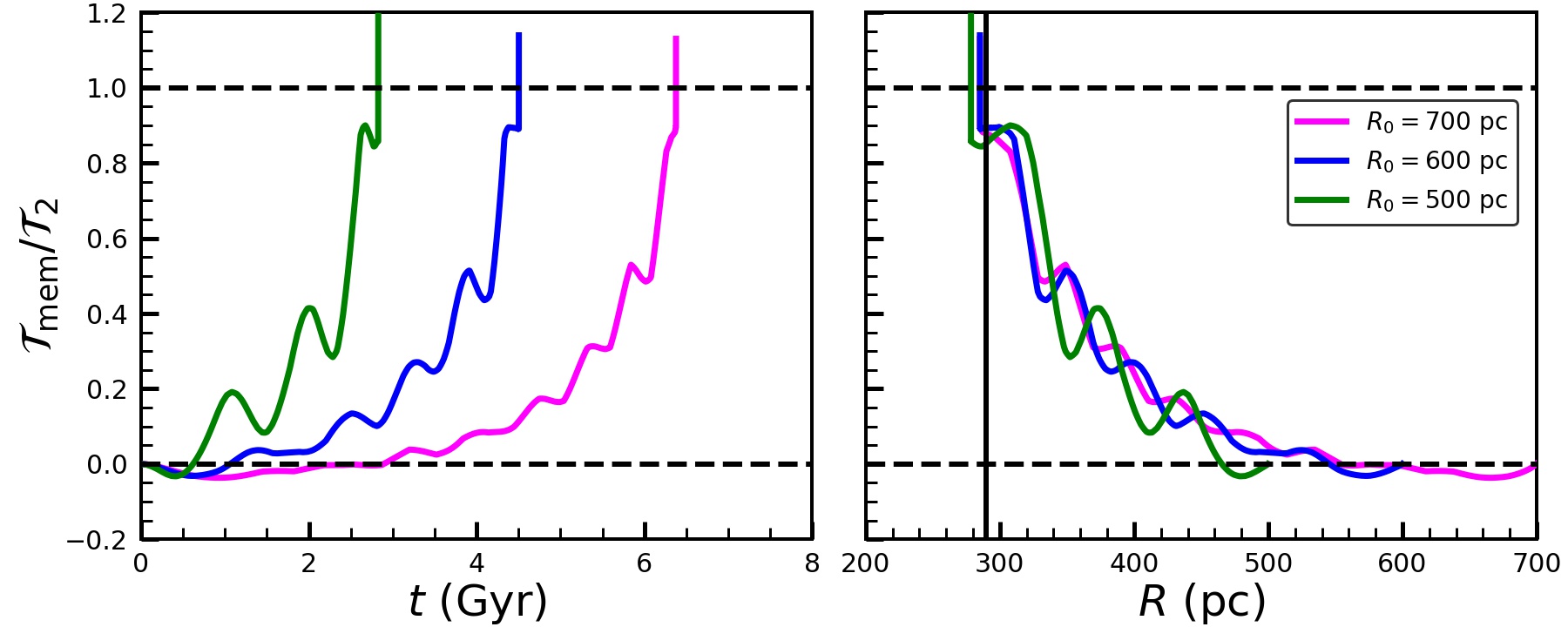}
\caption{The memory torque $\calT_{\rm mem}$ normalized by the total torque $\calT_2$ and computed using the $10$ dominant $(m,l)$ modes shown in Fig.~\ref{fig:LBKiso}, for the orbital decay of a point perturber in our fiducial isochrone sphere. Left (right) panel plots  $\calT_{\rm mem}/\calT_2$ vs $t$ ($R$) for three different initial radii $R_0$ as indicated. Note that the memory torque is initially retarding and sub-dominant but gradually gains strength, while undergoing oscillations, until it dominates (causing the accelerated Super-Chandrasekhar infall) near the critical radius $\Rcrit$ (marked by the vertical black line in the right-hand panel), where it flips sign, making the total torque enhancing (dynamical buoyancy).}
\label{fig:Torque_mem}
\end{figure*}

The solid lines in Fig.~\ref{fig:R_vs_t} plot the resulting orbital decay tracks, $R(t)$, obtained for 6 different initial radii, $R_0 = 700\pc, 600\pc,..., 200\pc$, which bracket $\Rcrit$. For comparison, we also show for each case the orbital decay track obtained using the LBK torque (dashed lines), and the standard Chandrasekhar formalism (dotted lines), which are obtained by solving equation~(\ref{dRdt_torque}) with $\calT_2 = \Tlbk$ and $\calT_2 = \bR \times \bF_{\rm DF}$, respectively. Here $\bF_{\rm DF}$ is given by equation~(\ref{aDF}) where we follow KS18 by setting $\ln{\Lambda}=\ln(R/a)$, with $a=10\pc$ the assumed scale radius for the perturber,\footnote{Since $a \ll b$, we are justified to treat the perturber as a point mass in the computation of the torque.} and properly compute $\rho(<v)$ from the isotropic distribution function of the isochrone sphere (equation~[\ref{f0}]).

When the initial radius of introduction, $R_0$, is large (compared to $R_{\rm crit}$), the orbital decay is characterized by four distinct phases of infall:
\begin{itemize}
    \item Phase I: Following the introduction into the system, the perturber falls in at a slightly slower rate than what is predicted by the LBK torque alone. This is because it takes time for the torque to build up and saturate to the asymptotic LBK value.
    
    \item Phase II: Once the transients have died out, and the torque has reached the steady LBK value, the infall rate of the perturber matches that predicted by the LBK torque.
    
    \item Phase III: As it approaches $\Rcrit$, the perturber starts to fall in at an accelerated pace, much faster than predicted by either the LBK torque or the standard Chandrasekhar formalism. This enhancement of the torque occurs only in the core region of the galaxy and is known as super-Chandrasekhar dynamical friction \citep[see e.g.,][]{Read.etal.06c, Goerdt.etal.10, Zelnikov.Kuskov.16}.
    
     \item Phase IV: Finally, the perturber reaches the stalling radius $R_{\rm st} \simeq \Rcrit$ about which it oscillates under the action of dynamical friction (retarding torque) outside and buoyancy (enhancing torque) within.
\end{itemize}

When the initial radius $R_0$ is close to $R_{\rm crit}$ the transients due to the self-consistent torque become more pronounced, in agreement with Fig.~\ref{fig:genLBK}. Introducing the perturber inside of the critical radius (i.e., $R_0 < R_{\rm crit}$) results in it being pushed out to $R_{\rm crit}$ (following initial transient oscillations).  Following \citet{Cole.etal.12} we refer to this as `dynamical buoyancy'.

This complicated behavior is in excellent, qualitative agreement with numerical $N$-body simulations, which have revealed how perturbers, upon approaching a central core, undergo accelerated super-Chandrasekhar friction, followed by a `kick-back' effect in which the perturber is pushed out again before it ultimately settles (stalls) at some radius, typically close to the core radius (see Section~\ref{sec:intro} for references).  

None of this is predicted by the standard Chandrasekhar formalism, according to which the perturber continues to sink all the way towards the center, albeit at a rate that becomes smaller towards the core. The latter owes to the fact that both $\rho(<v)$ and the Coulomb logarithm $\ln{\Lambda}=\ln(R/a)$ decrease with decreasing $R$. However, the resulting decline in the Chandrasekhar torque is insufficient to result in stalling. 

The LBK torque is more successful, in that it clearly predicts core stalling. As discussed in detail in KS18, the stalling is expected to occur at or near the `filtering radius', $R_\ast$, defined as the radius where the circular frequency of the perturber is equal to that of stars in the central core region \citep[see also][]{Read.etal.06c}. KS18 showed that the lower order modes, which otherwise exert a strong torque on the perturber, drop out of resonance, causing a significant reduction in the amplitude of the LBK torque (see also Fig.~\ref{fig:LBKiso}). This suppression of resonances arises from the fact that just outside of $R_\ast$, the circular frequency of the perturber is just a little bit lower than that of the core stars. Indeed, as shown by the dashed curves in Fig.~\ref{fig:R_vs_t}, based on the LBK torque one predicts that the infall stalls just outside of $R_\ast$ (indicated by the horizontal, brown line).  Note, though, that the LBK torque does neither predict a super-Chandrasekhar phase, nor dynamical buoyancy (the LBK torque is always retarding). In fact, introducing a perturber at $R_0 = 200 \pc < R_{\rm crit}$, the formalism based on the LBK torque predicts that it remains at that radius (see purple, dashed line in Fig.~\ref{fig:R_vs_t}),

Our formalism based on the self-consistent torque predicts a much richer dynamics, including dynamical buoyancy and super-Chandrasekhar infall. It also implies an explanation for core-stalling that is intriguingly different from that based on the LBK torque. Rather than resulting from a diminishing of the dynamical friction torque, core-stalling is an outcome of a balance between friction and buoyancy. All of this owes to the memory torque, which becomes dominant over the instantaneous torque close to $R_{\rm crit}$, and which causes the total torque to flip sign upon crossing $R_{\rm crit}$. This is illustrated in Fig.~\ref{fig:Torque_mem}, which plots $\calT_{\rm mem}/\calT_2$ as a function of time (left-hand panel) and radius (right-hand panel), respectively, for three different values of the initial radius $R_0$, as indicated. Initially, the torque is dominated by the instantaneous term, $\calT_{\rm inst}$, given by equation~(\ref{Torque_inst}). The memory torque slowly gains strength, while undergoing oscillations, and starts dominating when the perturber approaches $R_{\rm st}$ (indicated by the vertical black line in the right-hand panel). At this radius, $\calT_{\rm mem}$ (and thus also the total torque) flips sign and becomes enhancing, giving rise to dynamical buoyancy when $R < R_{\rm st}$, even though the instantaneous torque remains retarding.

\medskip

\subsection{Caveats and Outstanding Issues}
\label{sec:caveats}

Despite its success in reproducing previously unexplained aspects of dynamical friction observed in numerical simulations, in particular super-Chandrasekhar infall and dynamical buoyancy, the treatment of orbital decay based on the self-consistent torque presented above is subject to a few caveats.

First of all, we have ignored the time-evolution of $\Omega_\rmP$ (i.e., we assumed that $\Omega_\rmP = \overline{\Omega}_\rmP$). Although this is likely to be a reasonable approximation close to a constant density core, where $\Omega_\rmP(R)$ is nearly independent of radius, it remains to be seen how a proper treatment with a non-zero $\rmd\Omega_\rmP/\rmd t$ impacts the orbital decay. Unfortunately, since the temporal evolution of $\Omega_\rmP$, as quantified by $\overline{\Omega}_\rmP$ (equation~[\ref{barOmegaP}]), enters as an argument of the cosine and sine in the expressions for $\calJ_{1\boldell}(\bI,t)$ and $\calJ_{2\boldell}(\bI,t)$, respectively, numerically evaluating the corresponding integrals is non-trivial. 

Secondly, in deriving the expression for $\rmd R/\rmd t$ (equation~[\ref{dRdt_selfconst}]) we expanded $R(t)$ as a Taylor series that we truncated at first order. This is only valid as long as $\rmd^2 R/\rmd t^2$ is sufficiently small. Unfortunately, this is likely to be violated during the super-Chandrasekhar phase, when the rate of infall rapidly accelerates. This caveat, which is also responsible for the singular behavior at $\Rcrit$, can be overcome by using a higher-order truncation of the Taylor series, or by trying to directly solve the integro-differential equation~(\ref{dRdt}). We leave this as an exercise for future investigations. 

Finally, the entire formalism is based on perturbation theory, and therefore hinges on the assumption that the perturbation parameter $\vert \Phi_1^{\rm ext}/H_{0\rmJ}\vert$ is small. This assumption becomes questionable whenever the galaxy mass enclosed by the perturber, $M_\rmG(R)$, becomes comparable to the perturber mass. Unfortunately, in numerical simulations core stalling often happens at a radius at which $M_\rmG(R) \sim M_\rmP$ \citep[][]{Petts.etal.15, Petts.etal.16, DuttaChowdhury.etal.19}. In addition, when the perturber stalls at a fixed radius, the resonances no longer sweep by the stars fast enough to prevent non-linear perturbations from developing (i.e., one is no longer in what TW84 refer to as the `fast regime'). These non-linearities can even reverse the gradient of the distribution function near the resonances and contribute to an enhancing torque (which may counteract the retarding torque from the `fast' resonances and stall the infall) if the stars remain near-resonant for long enough ($\rmd\Omega_\rmP/\rmd t$ is slow enough), as is the case near the stalling radius. \cite{Sellwood.2006} finds such an effect in $N$-body simulations of a rotating bar-like perturbation in a spherical galaxy.

All of this suggests that a proper treatment of core stalling may not be possible with perturbation theory. In a follow-up paper (Banik \& van den Bosch, in prep.), we therefore examine dynamical friction, and core-stalling in particular, using a non-perturbative, orbit-based approach. This reveals a family of (perturbed) orbits that exert a coherent, enhancing torque, thus contributing to dynamical buoyancy. When the perturber approaches the central core region, the nature of the near-resonant orbits changes, due to a bifurcation of the inner Lagrange points, causing buoyancy to become dominant over friction. Hence, the non-perturbative, orbit-based approach lends support to our conclusion that central core regions manifest dynamical buoyancy, something that was first noticed in the numerical simulations by \citet{Cole.etal.12}.

\bigskip

\section{Conclusions}
\label{sec:concl}

Various approaches to describe dynamical friction in inhomogeneous systems have shown that it ultimately arises from a torque that has a non-zero contribution only from stars in resonance with the perturber. Ultimately, this notion that only the resonances contribute to the torque has its origin in the assumption that the orbital decay rate of the perturber is (secular approximation) and always has been (adiabatic approximation) very slow compared to the dynamical time of the host\footnote{This is similar to Bogoliubov's ansatz in plasma physics that the two-point correlation function relaxes much faster than the one-point distribution function.}. In the Hamiltonian perturbation theory of TW84 and KS18 the adiabatic approximation is enforced by multiplying the perturber potential by $\rme^{\gamma t}$ and taking the limit $\gamma \to 0$, while the secular approximation enters when the assumption is made that the orbital radius and circular frequency of the perturber are time-invariant over a dynamical time. This, in turn, implies that the equations of motion of both the perturber and the field particles are predominantly determined by the mean field, and thus characterized by slowly varying actions. Note that the same assumptions also underlie other approaches to dynamical friction in an inhomogeneous background, such as that based on the generalized Landau equation \citep[e.g.,][]{Chavanis.13} or the stochastic approach in action-angle space based on the fluctuation dissipation theorem \citep[e.g.,][]{Fouvry.Bar-Or.18}.

The secular and adiabatic approximations are justified when the mass of the perturber is sufficiently small. In that case, the dynamical friction time is much longer than the dynamical time. However, dynamical friction is mainly of astrophysical interest if the friction time is shorter than the Hubble time, which typically implies a perturber mass $M_\rmP$ in excess of 1-10 percent of the host mass. For such massive perturbers the dynamical friction time is no longer well separated from the dynamical time, and the secular and adiabatic assumptions are no longer justified. This breakdown is especially acute in the case of super-Chandrasekhar dynamical friction observed in numerical simulations when a massive perturber approaches a constant density core.

In this paper we have examined implications of relaxing the adiabatic and secular assumptions. Using Hamiltonian perturbation theory similar to KS18, but without taking the limit $\gamma \to 0$ and without the assumption that the response density builds up on the same time scale as that on which the perturber is introduced, we first relaxed the adiabatic approximation and derived an expression for the generalized LBK torque (equation~[\ref{Torque_gen}]). This differs from the standard LBK torque in that it depends on the growth rate $\gamma$ and has contribution from all orbits, resonant and non-resonant. Taking the adiabatic limit $\gamma\to 0$, i.e., assuming an extremely slow growth of the perturber potential, we recovered the LBK torque with a non-zero contribution only from the pure resonances. The opposite limit, $\gamma\to \infty$, corresponding to an instantaneous introduction of the perturber as typically done in idealized numerical simulations, leads to a time-dependent torque with a non-zero contribution from the near-resonant orbits along with the purely resonant ones. This `instantaneous' torque builds up linearly with time before undergoing oscillations (`transients') about the LBK value. Over time these oscillations damp out, and in the long-term the generalized torque reduces to the LBK torque. This behavior is analogous to how a forced, damped oscillator undergoes transients before settling to a steady-state solution in which the frequency of the response matches the driving frequency. The main difference is that here the damping is due to phase mixing of the responses from the individual orbits (each with its own frequencies), rather than due to some dissipative processes. The time-scale of relevance here is the time-scale on which the transients damp away, which is proportional to the dynamic range in orbital frequencies of the field particles that make up the host. Typically this range is sufficiently large and phase mixing is very efficient, causing the generalized LBK torque to quickly transition to the LBK torque. This justifies the standard treatments of dynamical friction, in which only the resonances contribute, even when the adiabatic approximation is not necessarily justified. However, there is one important exception, which is the case when the perturber is introduced close to a central constant-density core. Here the dynamic range of frequencies is drastically suppressed, causing large transient oscillations that can take many orbital periods of the perturber to phase mix away (see Fig.~\ref{fig:genLBK}).

Although the generalized LBK torque gives useful insight as to how transients that result from a non-adiabatic introduction of the perturber phase mix away, it is still based on the unphysical ansatz that the perturber grows its mass exponentially over time, on a characteristic time $\tau_{\rm grow} = 1/\gamma$, while remaining at a fixed host-centric radius, $R$. This time invariance of $R$ is a manifestation of the secular approximation. In order to improve on this, we next computed the torque in a self-consistent manner, in which we retained the information about the time dependence of the potential and circular frequency of the perturber throughout the entire evolution of the perturber's orbital radius, $R(t)$ (we relaxed the secular approximation). This self-consistent torque differs from the (generalized) LBK torque in that the instantaneous circular frequency of the perturber is replaced by its time-averaged value, and that it includes a convolution term that embodies the temporal correlation of the perturber potential. As a consequence, the self-consistent torque always has a non-zero contribution from the near-resonant orbits, and depends on the entire infall history, $R(t)$, which in turn is dictated by the torque itself. A proper description of dynamical friction thus requires solving an integro-differential equation for $R(t)$ (equation~[\ref{dRdt}]).

While solving this equation in full generality is highly non-trivial, we obtained some valuable insight by Taylor expanding $R(t)$ and truncating it at first order. This is valid as long as  $\rmd^2 R/\rmd t^2$ is sufficiently small, i.e., the rate of infall, $\rmd R/\rmd t$ varies slowly. If, in addition, we assume that the time-dependence of the perturber's frequency is small, which is a valid assumption at or near a central core region, we can write the self-consistent torque as a sum of two terms, the instantaneous torque, which depends on $R(t)$, and a memory torque, which is proportional to $\rmd R/\rmd t$ besides having an $R$ dependence. We used this simplified form of the self-consistent torque to evolve the radius $R(t)$ of a point mass perturber in an isochrone galaxy (which has a central core). We found that the infall of the perturber occurs in four subsequent phases: (i) sub-LBK infall during the initial (linear) build-up of the torque, (ii) infall at the LBK rate as the instantaneous torque asymptotes to the LBK torque, (iii) accelerated super-Chandrasekhar infall due to a destabilizing effect of the memory torque, and (iv) kick-back of the perturber from within a critical radius $\Rcrit$ due to buoyant effects followed by stalling at that radius. The instantaneous torque dominates the early phase of the infall while the memory torque becomes dominant near the critical radius. It is responsible for the super-Chandrasekhar infall and flips sign at $\Rcrit$, causing the total torque to become enhancing for $R<\Rcrit$. When the perturber is introduced inside of $\Rcrit$, it is consequently pushed out (dynamical buoyancy) to $\Rcrit$ by this enhancing memory torque.

These phenomena of super-Chandrasekhar infall followed by kick-back and core-stalling, as well as dynamical buoyancy inside central core regions, have been observed in numerous $N$-body simulations \citep[][]{Read.etal.06c, Goerdt.etal.10, Inoue.11, Cole.etal.12, DuttaChowdhury.etal.19}, but have thus far eluded a proper explanation \citep[but see][for some phenomenological explanations]{Read.etal.06c, Petts.etal.15, Petts.etal.16, Zelnikov.Kuskov.16}. Although KS18 had shown that the LBK torque strongly diminishes as one approaches a core, which they advocated as an explanation for core stalling, they were unable to explain either super-Chandrasekhar infall or dynamical buoyancy. Based on our results, we argue that core-stalling is ultimately a consequence of a subtle balance between dynamical friction (retarding torque) and buoyancy (enhancing torque), which is preceded by a phase of super-Chandrasekhar friction caused by the destabilizing effect of the memory torque that depends on the past infall history.

Finally, while wrapping up this paper, we became aware of an unpublished study by M. Weinberg \citep[][]{Weinberg.04}, in which they also point out the problematic nature of the `time-asymptotic limit' (i.e., taking $\gamma \to 0$) used to derive the LBK torque. Using Hamiltonian perturbation theory similar to what is presented here, but using Laplace transforms rather than Green's functions to solve for the response, they obtain a time-dependent torque (equation~[14] in their paper) that is identical to our self-consistent torque of equation~(\ref{Torque_SC}), except that it doesn't explicitly account for a time-dependence of the perturber frequency. They then proceed to examine how the time-dependent torque differs from the LBK torque for the examples of a slowing bar and a decaying satellite. In the latter case, rather than calculating the orbital decay of the satellite self-consistently, as done here, they first compute the orbital decay $R(t)$ using the local Chandrasekhar formula, which is then substituted in the expression for the time-dependent torque. In agreement with our results, they show that massive perturbers, which decay rapidly, are significantly impacted by transients that are not accounted for in the LBK torque.

\section*{Acknowledgments}

We are thankful to the anonymous referee for insightful comments and suggestions. We are grateful to Martin Weinberg for an enlightening discussion about the role of non-resonant orbits in dynamical friction, which ultimately set us on the path to explore a self-consistent treatment not knowing that he had worked on this topic before. FvdB is supported by the National Aeronautics and Space Administration through Grant Nos. 17-ATP17-0028 and 19-ATP19-0059 issued as part of the Astrophysics Theory Program.

\bigskip

%%%%%%%%%%%%%%%%%
% Bibliography
%%%%%%%%%%%%%%%%%

\bibliography{references_vdb}{}
\bibliographystyle{aasjournal}

%%%%%%%%%%%%%%%%% APPENDICES %%%%%%%%%%%%%%%%%%%%%

\appendix
\section{The isochrone sphere}
\label{app:model}

All specific examples presented in this paper correspond to a point mass perturber, with mass $M_\rmP$, moving on a circular orbit in an isotropic isochrone sphere, whose potential and density are given by equations~(\ref{IsoPot}) and~(\ref{IsoDens}), respectively. In addition, the distribution function of the (unperturbed) isotropic isochrone sphere of mass $M_\rmG$ is given by
\begin{align}
f_0(\varepsilon) = \frac{M_\rmG}{\sqrt{2}{(2\pi)}^3 {(G M_\rmG b)}^{3/2}} \frac{\sqrt{\varepsilon}}{{\left[2(1-\varepsilon)\right]}^4} \left[27-66\varepsilon+320\varepsilon^2 - 240\varepsilon^3 + 64\varepsilon^4 + 3(16\varepsilon^2 + 28\varepsilon - 9) \frac{\arcsin \sqrt{\varepsilon}}{\sqrt{\varepsilon(1-\varepsilon)}} \right],
\label{f0}
\end{align}
\citep[e.g.,][]{Binney.Tremaine.08}, where $\varepsilon = -E_0\, b/G M_\rmG$ and covers the range $0<\varepsilon\leq 1/2$. 

In absence of the perturber, the orbits of the field particles are characterized by four isolating integrals of motion: the energy $H_0$ and the three actions $(I_r,L,L_z)$. Following KS18 we make a canonical transformation from $(I_r,L,L_z)$ to $(I,L,L_z)$ where $I$ is given by $2 I_r + L$, with $0\leq L \leq I$ and $-L\leq L_z\leq L$. In terms of $I$ and $L$, the orbital energy per unit mass is given by
\begin{align}
E_0(I,L) = -\frac{2{\left(G M_\rmG\right)}^2}{{\left[I+\sqrt{I^2_\rmb+L^2}\right]}^2}\,,
\label{E0}
\end{align}
where $I_b \equiv 2\sqrt{G M_\rmG b}$. While $E_0$ is conserved in the inertial frame, in the co-rotating perturbed frame the conserved quantity is the Jacobi Hamiltonian, given by
\begin{align}
H_{\rmJ 0}(I,L,L_z) = E_0(I,L) - \Omega_\rmP L_z\,.
\label{HJ0}
\end{align}

Corresponding to the actions are the conjugate angles $(w,g,h)$, whose corresponding frequencies are given by
\begin{align}
&\Omega_w(I,L) = \frac{\partial H_{\rmJ 0}}{\partial I} = \frac{\Omega_r}{2} = \frac{4{\left(G M_\rmG\right)}^2}{{\left[I+\sqrt{I^2_\rmb+L^2}\right]}^3}, \nonumber \\
&\Omega_g(I,L) = \frac{\partial H_{\rmJ 0}}{\partial L} = \Omega_\psi-\frac{\Omega_r}{2} = \frac{L}{\sqrt{I^2_\rmb+L^2}} \Omega_w(I,L), \nonumber \\
&\Omega_h(I,L) = \frac{\partial H_{\rmJ 0}}{\partial L_z} = -\Omega_\rmP.
\label{frequencies}
\end{align}
Here $\Omega_g$ is the frequency of periapse precession. $\Omega_r$ and $\Omega_\psi$ are the radial and angular frequencies in the orbital plane, which can be expressed in terms of the actions $I$ and $L$ as
\begin{align}
&\Omega_r (I,L) = \frac{8{\left(G M_\rmG\right)}^2}{{\left[I+\sqrt{I^2_\rmb+L^2}\right]}^3}, \nonumber \\
&\Omega_\psi (I,L) = \frac{\Omega_r}{2} \left(1+\frac{L}{\sqrt{I^2_\rmb+L^2}}\right)\,.
\end{align}
The fact that all these (unperturbed) frequencies can be expressed as simple algebraic functions is what makes the isochrone potential ideal for an analytical exploration of core-stalling. 

Following KS18, we ignore the torque from the stars outside of the core, which allows us to truncate the integration over $I$ at a maximum value $I_{\rm max} \ll I_b$. We follow KS18 and adopt $I_{\rm max}=0.1\,I_b$. Under this approximation the expressions for the frequencies can be simplified as follows
\begin{align}
&\Omega_w \approx \Omega_b\left(1-3\frac{I}{I_b}\right), \nonumber \\
&\Omega_g \approx \Omega_b \frac{L}{I_b},
\label{Omega_wg}
\end{align}
where $\Omega_b = 0.5\sqrt{G M_\rmG/b^3}$ is the central frequency of the galaxy.

Substituting the above expressions for the frequencies, we have the following expression for the resonance angle
\begin{align}
\ell_k\Omega_k = n\,\Omega_w + l\,\Omega_g - m\,\Omega_\rmP = s\,\Omega_b - m\,\Omega_\rmP\,,
\end{align}
where 
\begin{align}
s = s(I,L) \equiv \left[n \left(1 - 3\frac{I}{I_b}\right) + l\frac{L}{I_b}\right]\,.
\end{align}
KS18 find that the co-rotation resonances with $n=m$ exert much stronger torque than the ones with $n\neq m$; therefore we shall only study co-rotation modes in this paper. 

Substituting the above expressions in equation~(\ref{Torque_inst}), we arrive at the following form for the instantaneous torque for the $(m,l,m)$ mode
\begin{align}
\calT_{2,ml} &= 16\pi^3 m \Omega_b \int_0^{I_{\rm max}} \rmd I \int_0^I \rmd L \,\dfrac{\sin{\left[s\,\Omega_b-m\,\Omega_\rmP\right] t}}{s\,\Omega_b-m\,\Omega_\rmP} \,s(I,L) \,\frac{\partial f_0}{\partial E_0}\, P_{mlm}(I,L),
\label{Torque_inst_isochrone}
\end{align}
where $P_{mlm}(I,L)$ is given by
\begin{align}
P_{mlm}(I,L)=\int_{-L}^{L} \rmd L_z {\left|\hat{\Phi}'_{mlm}(I,L,L_z)\right|}^2.
\end{align}
We compute the Fourier coefficients $\hat{\Phi}'_{mlm}(I,L,L_\rmz)$ using the analytical expressions given in Appendix A of KS18. 

The corresponding LBK torque is given by
\begin{align}
\calT_{2,ml}^{\rm LBK} &= 16\pi^4 m^2\, \Omega_\rmP \int_0^{I_{\rm max}} \rmd I \int_0^I \rmd L \,\delta\left[s\,\Omega_b - m\,\Omega_\rmP\right] \frac{\partial f_0}{\partial E_0}\, P_{mlm}(I,L).
\label{Torque_LBK_isochrone}
\end{align}

\bigskip

%%%%%%%%%%%%%%%%%%%%%%%%%%%%%%%%%%%%%%%%%%%%%%%%%%
% Don't change these lines
%\bsp	% typesetting comment
\label{lastpage}



\end{document}